\documentclass[pdflatex,sn-mathphys-num]{sn-jnl}

\usepackage{graphicx}%
\usepackage{multirow}%
\usepackage{amsmath,amssymb,amsfonts}%
\usepackage{amsthm}%
\usepackage{mathrsfs}%
\usepackage[title]{appendix}%
\usepackage{xcolor}%
\usepackage{textcomp}%
\usepackage{manyfoot}%
\usepackage{booktabs}%
\usepackage{algorithm}%
\usepackage{algorithmicx}%
\usepackage{algpseudocode}%
\usepackage{listings}%
\usepackage{lineno}

\theoremstyle{thmstyleone}%
\theoremstyle{thmstyletwo}%

\theoremstyle{thmstylethree}%

\begin{document}

\title[Article Title]{Solar Magnetic Configuration Control over Radiation Belt Electrons}


\author*[1]{\fnm{Ahmad} \sur{Lalti}}\email{ahmad.lalti@northumbria.ac.uk, ahmad.lalti@gmail.com}
\author[1]{\fnm{Jonathan} \sur{Rae}}\email{jonathan.rae@northumbria.ac.uk}
\author[1]{\fnm{Clare} \sur{Watt}}\email{clare.watt@northumbria.ac.uk}
\author[1,2,3]{\fnm{Stephanie} \sur{Yardley}}\email{steph.yardley@northumbria.ac.uk}
\author[4]{\fnm{Savvas} \sur{Raptis}\email{savvas.raptis@jhuapl.edu}}
\affil*[1]{\orgname{Northumbria University}, \orgaddress{\street{Ellison PI}, \city{Newcastle Upon Tyne}, \postcode{NE1 8ST}, \state{England}, \country{United Kingdom}}}
\affil[2]{\orgname{University College London, Mullard Space Science Laboratory}, \orgaddress{\street{Holmbury St. Mary}}, \city{Dorking}, \postcode{RH5 6NT, UK}, \state{England}, \country{United Kingdom}}
\affil[3]{\orgname{Donostia International Physics Center (DIPC)}, \orgaddress{\street{Paseo Manuel de Lardizabal 4}}, \postcode{20018}, \state{San Sebasti{\'a}n}, \country{Spain}}
\affil[4]{\orgdiv{Johns Hopkins University}, \orgname{ Applied Physics Laboratory}, \orgaddress{\city{Laurel},  \state{MD}, \country{USA}}}


\abstract{
Earth is surrounded by two highly dynamic concentric belts of particle radiation. The outer radiation belt exhibits coherent variability at solar-cycle, seasonal, and 27-day (Carrington) timescales. While the solar-cycle and Carrington variations have been attributed to the recurrence of coronal hole solar wind, the seasonal variation has long been explained by local geometric effects that modulate the solar wind-magnetosphere coupling. Here, we challenge this paradigm by showing that the periodic recurrence of coronal hole solar wind likewise drives the seasonal variation. We further demonstrate that the Alfv\'enic nature of this solar wind is a key driver of the observed electron flux enhancement in the outer radiation belt. These findings provide a unifying framework linking solar magnetic topology, solar wind properties, and magnetospheric dynamics across multiple timescales at Earth and beyond.


}

\maketitle

\section{Introduction}\label{sec1}
One of the most striking discoveries made with the advent of the space age is the existence of belts of highly relativistic charged particles trapped within Earth's magnetic field \cite{van1958observation,vernov1960investigations,stepanova2024regarding}. Earth is surrounded by two concentric belts of energetic charged particles, an inner belt made primarily of protons and electrons, and an outer belt of relativistic electrons \cite{li2019Earth}. Such relativistic particles have adverse effects on satellites, particularly in Geostationary Earth Orbit (GEO) and Medium Earth Orbit (MEO) \cite{hands2018radiation,eastwood2017economic,wrenn1995conclusive}. To design and operate satellites safely in such a hostile environment, we must understand the dynamics of Earth's radiation belts and accurately model and predict their behavior. Radiation belts are not unique to Earth; they have been observed at other magnetized celestial bodies in the solar system \cite{mauk2010electron} and beyond \cite{climent2023evidence,kao2023resolved,turner2026universal}. This makes Earth's radiation belts a natural laboratory for understanding radiation belts across the cosmos.

The Earth's inner belt is dynamically stable and exhibits little temporal variation, with charged particles trapped for months and years \cite{claudepierre2019revised,li2019Earth}. In contrast, the outer electron radiation belt exhibits highly complex and nonlinear dynamics across various timescales, which are mainly driven by the interaction between the solar wind and Earth's magnetosphere \cite{turner2012explaining,reeves2013electron,reeves2013long,jaynes2015source}. This nonlinear behavior makes it challenging to model and accurately forecast the flux of electrons in the outer radiation belt.

The outer electron radiation belt exhibits two types of variability. The first is driven by aperiodic solar wind transient structures such as interplanetary coronal mass ejections (ICMEs) \cite{echer2008interplanetary,yermolaev2012geoeffectiveness,murphy2018global}. The second is a more coherent, periodic variability, which will be the focus of this study.

On solar cycle timescales, observations show that the peak electron flux in the outer radiation belt is reached during the declining phase \cite{baker2001global,yermolaev2012geoeffectiveness,tsurutani1996interplanetary,tsurutani2006corotating,hajra2014relativistic}.
On shorter timescales, of the order of the Carrington period ($\sim 27$ days) and below, periodic flux enhancements are observed during what is known as high intensity long-duration continuous auroral electrojet (AE) activity (HILDCAA) events \cite{tsurutani1987cause,tsurutani1995interplanetary,tsurutani2006corotating,hajra2024ultra}. 

Both periodicities have been attributed to the recurrent interaction of Earth's magnetosphere with stream interaction regions (SIRs) and the subsequent high-speed streams (HSS) emanating from coronal holes \cite{tsurutani1996interplanetary,tsurutani2006corotating,hajra2022corotating}. Many studies attribute the geoeffectiveness of HSS to large-amplitude Alfv\'en waves embedded in the coronal hole interplanetary magnetic field (IMF), which enhance coupling with the magnetosphere \cite[and references therein]{tsurutani1995interplanetary,lyons2005solar,lee2006repetitive,tsurutani2006corotating,hajra2014relativistic,ala2024impact}. Additionally, HSS with predominantly southward IMF Bz have been found to produce larger flux enhancements than those with northward Bz \cite{miyoshi2013high}. Other studies instead argue that the high flow speeds within HSS favor large-scale magnetopause instabilities and the subsequent generation of Ultra-Low-Frequency (ULF) waves in the inner magnetosphere \cite[and references therein]{mathie2001solar,mann2004correlations,kavosi2015ubiquity,georgiou2018ultralow,murphy2018determining}. The two pathways (Alfv\'enicity and speed) are not mutually exclusive since they may act in concert to accelerate electrons in the outer radiation belt. However, the relative contribution of each mechanism remains to be established. Here we present direct observational evidence that solar wind Alfv\'enicity is a key driver of electron energization in the outer radiation belt. We further show that high solar wind speed is not a necessary condition for this energization.

On intermediate timescales, the flux of electrons in the outer radiation belt has been observed to oscillate with a seasonal (6-month) period \cite{paulikas1979effects,baker1999equinoctial,li2001long,hajra2020seasonal}. Three main mechanisms have been proposed to explain this seasonal variation \cite{lockwood2016origins,lockwood2020semi}. The first is the Russell-McPherron (RM) effect \cite{russell1973semiannual}, in which the tilt between Earth's equatorial plane and the ecliptic produces a seasonal modulation of the IMF $B_z$. Near the equinoxes and solstices, $B_z$ reaches its largest and smallest values, respectively, leading to corresponding variations in magnetospheric energy input and, ultimately, in outer-belt electron flux. The second proposed mechanism is termed the equinoctial effect \cite{boller1970kelvin}, wherein the angle between the solar wind flow and the Earth's dipole moment controls geomagnetic activity. Near the equinoxes, it is hypothesized that the magnetosphere is more prone to large-scale instabilities that drive elevated geomagnetic activity and, in turn, enhance electron flux. The third mechanism is the axial effect \cite{cortie1912sun}, which attributes the seasonal variation to Earth's heliographic latitude, modulating the likelihood of encountering high-speed streams from coronal holes as Earth moves in and out of the heliospheric streamer belt. Many recent studies have presented evidence in favor of the RM or equinoctial effects as the dominant driver of the seasonal variation \cite{mcpherron2009role,katsavrias2021semi,cliver2000mountains,li2001long,baker1999equinoctial,kanekal2010seasonal}, while the axial effect has received comparatively little support.

Here, we challenge the current paradigm that the seasonal variation is driven by local geometric effects (the RM and equinoctial effects) and show that it is, in fact, driven by the recurrence of coronal hole solar wind (the axial effect). Combined with our direct observational evidence linking Alfv\'enic solar wind to periodic electron energization, these results establish a unifying framework connecting solar magnetic topology, solar-wind structure, and magnetospheric dynamics across multiple timescales, with implications extending from Earth to other magnetized planets in the solar system and beyond.

\section{Results}\label{sec2}

\begin{figure}[htbp!]
    \centering
    \includegraphics[width=1\linewidth]{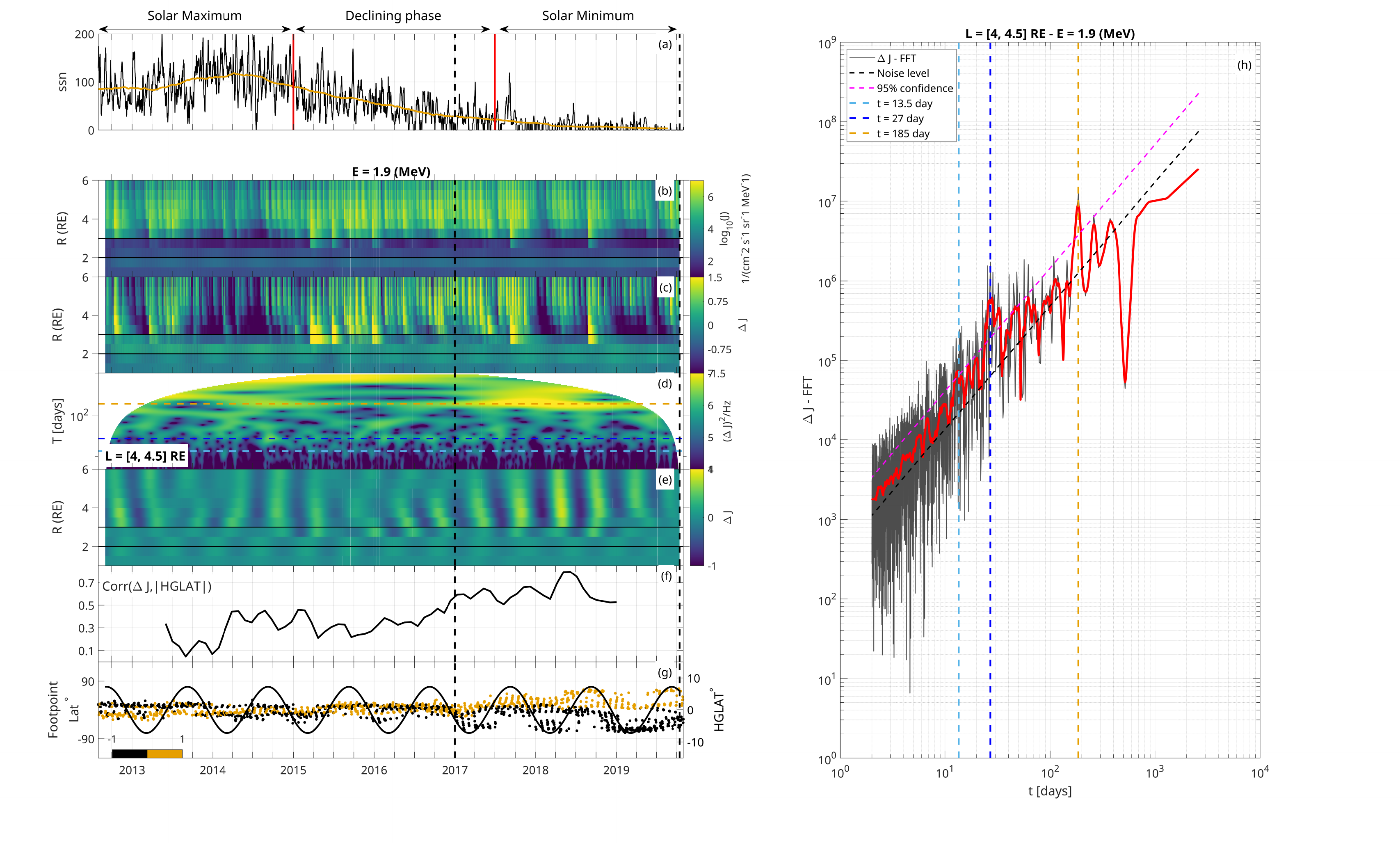}
    \caption{Seven years of Van Allen Probes (VAPs) electron flux data illustrating coherent and non-coherent variability across multiple timescales. Panel a shows the sunspot number at 1-hour resolution (black) and smoothed with a 1-year running average (orange). Panel b shows the pitch angle-integrated flux in log scale, $log_{10}(J)$, as a function of the geocentric radial distance R and time for the energy $E = 1.9$ MeV. For the same energy, panel c shows the normalized flux $\Delta J$ as a function of $R$ and time. Panel d shows the wavelet transform for a cut of $\Delta J$ taken at $R = [4, \; 4.5] \; R_E$. Panel e shows $\Delta J(R,t, E = 1.9 \; MeV)$ band pass filtered around the period of 185 days (filter interval $[150,\; 200]$ days). The horizontal black lines in panels b,c, and e highlight the average slot region in the interval $R = [2, \;3] \; R_E$. Panel f shows the cross correlation between $\Delta J$ and the absolute value of the heliographic latitude $|HGLAT|$ calculated in a sliding window of length 1.5 years and a step of 1 month. Panel g shows on the left axis the latitude at the solar surface of the footpoint of the solar wind reaching Earth (see Methods). Dots in black/orange represent the negative/positive magnetic field orientation. The right axis of panel g and the corresponding black line show the heliographic latitude (HGLAT) of Earth throughout the seven-year period. Panel h shows in grey the Fourier transform of a $\Delta J$ cut taken at $R = \left[4, \; 4.5\right] \; R_E$ and in red the same Fourier spectrum but filtered to remove the noise and highlight the peaks. The black and magenta dashed lines are the fit of the background noise in the Fourier spectrum and the $95\%$ confidence range to highlight the statistically relevant peaks. Dashed orange, blue, and cyan lines in panels d and h mark the three dominant periodicities: $\sim$185 days ($\sim 6$ months), 27 days, and 13.5 days.}
    \label{fig_overview}
\end{figure}

The Van Allen Probes (VAPs) mission \cite{TheVanAllenProbe2014fox}, was a dual satellite mission flown in an elliptical orbit with a focus on exploring the physics of Earth's radiation belts. Equipped with high resolution fields and comprehensive particle instrument suite, the VAPs were launched in August 2012 and decommissioned in December 2019, operating from the maximum of solar cycle 24 until the minimum of cycle 24 and the start of cycle 25, shown by the sunspot number in Fig. \ref{fig_overview}a.

In Fig. \ref{fig_overview}, we show seven years of the pitch-angle-integrated, $1.9$ MeV relativistic differential electron flux in log scale, $log_{10}(J)$ (panel b), and the flux normalized to its seven-year median, $\Delta J$ (panel c), measured by the Relativistic Electron Proton Telescope (REPT) instrument \cite{baker2013relativistic} from the combined VAP A and B (see Methods). Both the raw flux $J$ and the normalized flux $\Delta J$ exhibit dynamic variability at multiple timescales. To investigate the different periodicities, we perform a Fourier analysis on $\Delta J$ (panel h; see Methods), which clearly shows that the outer radiation belt exhibits coherent variability at timescales corresponding to 185 days (seasonal variation), the 27-day Carrington period, and its harmonics as noted by the vertical dashed lines and the corresponding peaks above the 95 $\%$ confidence level (magenta dashed line). A wavelet transform of $\Delta J$ (panel d) further reveals that these periodicities undergo clear dynamical change throughout the solar cycle, which we discuss below.

\subsection{The driver of the seasonal radiation belt enhancements}
\label{seasonal_variation}
We first examine the largest time-scale variation by analyzing the seven years of $\Delta J$ (Fig. \ref{fig_overview}c-d). The behavior differs clearly across the solar maximum (August 2012 - December 2014), the declining phase (January 2015 - June 2017), and the solar minimum (July 2017 - December 2019). During solar maximum, electron dynamics are dominated by sporadic flux enhancements driven by transient solar-wind structures such as ICMEs. On average, the flux remains 1-2 orders of magnitude below the seven-year median (most clearly in 2014). Periodic Carrington-scale enhancements are observed between February and November 2013 and again between August and December 2014 (see Extended data Fig \ref{fig_carrington}c,d). During the declining phase, the flux is, on average, 1-2 orders of magnitude above the seven-year median, with recurrent Carrington-period flux enhancements. By solar minimum, two overlapping periodicities emerge: a six-month seasonal enhancement and a modulation at the Carrington period (Fig. \ref{fig_overview}d, g).

This behavior is directly linked to the evolution of the solar magnetic field configuration throughout the solar cycle. At solar maximum, the Sun's dipole moment vanishes as it flips polarity, and higher order moments dominate the magnetic field \cite{woo2002origin,livshits2006variations}. As a result, the solar wind reaching Earth is dominated by slow streamer-belt wind, with fewer coronal hole fast wind contributions \cite{owens2013heliospheric,d2021solar}. However, the Sun is highly active during this phase, and the ICME rate reaches its peak \cite{richardson2012solar}, which becomes the main driver of electron flux enhancement in the outer radiation belt. The prolonged low fluxes during this phase, 1-2 orders of magnitude below the seven-year median, most likely reflect the scarcity of efficient energization drivers combined with the continuous operation of loss mechanisms \cite{millan2007review}. Occasionally, a persistent equatorial coronal hole forms, producing recurrent SIRs that enhance flux, as observed in 2013 and late 2014 (see Fig. \ref{fig_SDO} in the Extended data).

During the declining phase, the solar magnetic field restructures, with the dipole moment tilting relative to the solar equator and the ecliptic plane \cite{livshits2006variations}. This tilt allows polar coronal holes to reach lower latitudes. As a result, Earth experiences coronal hole solar wind once or twice each Carrington rotation, depending on the solar magnetic configuration. This extended and recurrent exposure saturates the outer radiation belt, producing its highest flux levels, as shown in Fig. \ref{fig_overview}.

To understand the behavior of the radiation belt during solar minimum, we apply a narrow band-pass filter to $\Delta J$ centered on the six-month periodicity (Fig. \ref{fig_overview}e). The coherence of these oscillations is lowest during solar maximum and gradually increases throughout the declining phase, reaching its peak at solar minimum (see signal power evolution around the orange dashed line in Fig. \ref{fig_overview}d and the amplitude of the filtered signal in panel e). This observation is inconsistent with both the RM and equinoctial effects. Both the RM and equinoctial effects are purely geometric mechanisms and are independent of solar wind type. If either mechanism were the primary driver, the seasonal variation should persist throughout the solar cycle, with only its amplitude modulated. Instead, we observe the opposite: the seasonal variation is weak or absent during solar maximum and much of the declining phase (Fig. \ref{fig_overview}d-e). The brief periods of weak coherence near 2012 and late 2014 coincide with intervals where multiple ICMEs and recurrent SIRs occur together, which can introduce mild coherence into the filtered signal.

On the other hand, a central assumption of the axial effect is that the solar dipole moment is almost aligned with the solar rotation axis \cite{cortie1912sun,lockwood2016origins}. This magnetic configuration is expected primarily during solar minimum \cite{woo2002origin,livshits2006variations}. Under these conditions, during solstices at low heliographic latitude, Earth crosses the heliospheric current sheet (HCS) and the solar wind reaching Earth is dominated by streamer belt (SB) wind. As the equinoxes approach, the absolute value of the heliographic latitude $|HGLAT|$ increases, exposing Earth to high-speed streams emanating from polar coronal holes, whose enhanced geoeffectiveness drives the seasonal enhancement of electron flux. Therefore, if the axial effect is valid, the seasonal variation should peak during solar minimum when the solar magnetic configuration is most favorable to drive this modulation, which is consistent with our observation in Fig. \ref{fig_overview}d,e.

Additionally, the axial effect predicts a correlation between the electron flux enhancement and $|HGLAT|$. It also predicts that the solar wind should alternately connect to the northern and southern polar coronal holes as $|HGLAT|$ peaks, oscillating with a six-month periodicity. We test those predictions by performing a Spearman cross-correlation analysis between $|HGLAT|$ and $\Delta J$, using a 1.5-year sliding window with a 1-month step (Fig. \ref{fig_overview}f; see methods section). This correlation peaks as we approach solar minimum (between the vertical dashed lines), with values $>0.6$ occurring when $\Delta J$ is time lagged by 30-45 days relative to $|HGLAT|$.
We then track the latitude of the solar footpoint and the polarity of the solar magnetic field (Fig. \ref{fig_overview}g; see methods section). We find a clear correlation between the coherence of the seasonal flux variation (Fig \ref{fig_overview}d, e) and the latitude of the solar source of the solar wind (Fig. \ref{fig_overview}g), see section \ref{correlation_lat_seasonal} in the Extended Data for a quantitative correlation analysis. During solar maximum and the declining phase, when seasonal variation is least coherent (Fig. \ref{fig_overview}d), the solar wind reaching Earth is primarily equatorial with footpoints constrained between $\sim -30^\circ$ and $\sim 30^\circ$. As solar minimum approaches, solar wind source footpoints move toward higher latitudes, and the solar magnetic polarity switches with each cycle of $|HGLAT|$ (Fig. \ref{fig_overview}g), indicating that Earth is periodically exposed to the Northern and Southern polar coronal holes on a semi-annual basis.

Finally, previous studies arguing for the Russell-McPherron (RM) effect \cite{mcpherron2009role,katsavrias2021semi}, the equinoctial effect \cite{cliver2000mountains,li2001long}, or both \cite{baker1999equinoctial,kanekal2010seasonal} as the driver of the seasonal variation in the radiation belt fluxes relied on comparing the timing of flux peaks in observations with those predicted by each mechanism. However, this approach is inherently uncertain as the timing of relativistic electron flux peaks varies substantially from year to year \cite{kanekal2010seasonal,hajra2020seasonal}. This approach also assumes a near-instantaneous response of the radiation belts to changes in the driver. This assumption has yet to be tested, and seems unlikely given that the magnetosphere behaves as a nonlinearly driven circuit with both resistive and inductive components \cite{akasofu2013relationship}. 

Taken together, these considerations support the conclusion that the axial effect, whereby Earth is exposed to polar-coronal-hole solar wind near the equinoxes at maximum heliographic latitude, is the dominant driver of the seasonal variation, rather than Russell-McPherron or equinoctial mechanisms.

\subsection{The influence of solar wind source and its cross helicity on radiation belt flux}

In the previous section, we showed that the axial effect drives the strong seasonal variation in outer radiation belt fluxes. Earlier work demonstrated that periodic exposure to high-speed streams (HSS) from coronal holes produces solar-cycle and Carrington periodicities in the outer electron belt \cite{tsurutani1996interplanetary,tsurutani2006corotating,hajra2022corotating}. Since the axial effect also arises from periodic exposure to coronal hole solar wind, our results unify the mechanism responsible for coherent outer-belt variability across multiple timescales. This naturally raises a central question: what properties of coronal hole solar wind make it particularly effective at energizing the radiation belt electrons?

Coronal hole solar wind could energize radiation belt electrons through two pathways: its Alfv\'enic fluctuations and its elevated bulk flow speed.
On one hand, the high-amplitude Alfv\'en waves embedded in coronal hole solar wind generate recurrent extended intervals of strong southward magnetic fields ($B_z<0$), which drive enhanced substorm activity through magnetic reconnection with the northward magnetospheric field \cite{tsurutani1987cause,tsurutani2006corotating}. On the other hand, high solar wind speeds lead to large shearing at the magnetopause, which creates the condition for the growth of the Kelvin-Helmholtz instability \cite{kavosi2015ubiquity}. Such surface fluctuations excite ULF waves in the inner magnetosphere \cite{mathie2001solar} which themselves have been linked to electron energization \cite{georgiou2018ultralow,murphy2018determining}. The two pathways are not mutually exclusive, and solar wind speed and its Alfv\'enicity may act in concert to energize electrons.

\begin{figure}[htbp!]
    \centering
    \includegraphics[width=1\linewidth]{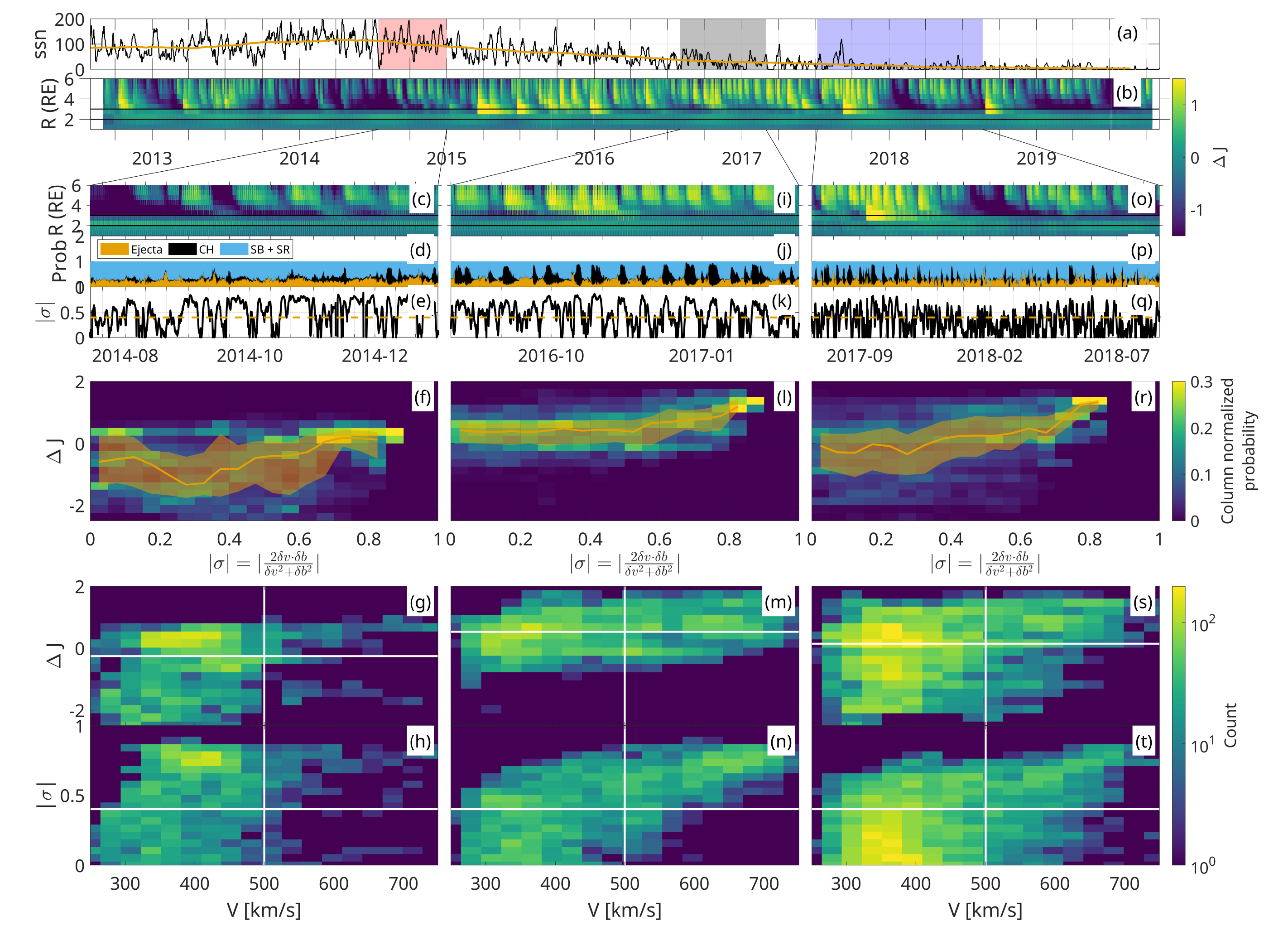}
    \caption{Three intervals at different phases of the solar cycle, highlighting the cause of the flux enhancement at Carrington periods. Panel a shows the sunspot number throughout the seven years of the VAPs mission. In black is the hourly sunspot number (ssn), and in orange is the ssn smoothed with a 1-year running average. Panel b is $\Delta J$ as a function of $R$ and time for the energy channel $E = 1.9$ MeV. Panels c-e, i-k, and o-q are a zoom-in on three different intervals representing solar maximum, the declining phase, and solar minimum, respectively. Panels c, i, and o plot $\Delta J$. d, j, and p are the probabilities that the solar wind is one of four different origins obtained using a classification technique \cite{xu2015new,camporeale2017classification}. Orange represents Ejecta, black coronal hole (CH), and blue represents the combination of streamer belt (SB) and sector reversal (SR). The dominant color at any given time indicates the most probable solar wind source. Panels e, k, and q are the absolute value of the normalized cross helicity ($|\sigma|$) as defined in the methods section. Panels f, l, and r show the column-normalized 2D probability distribution of $\Delta J$ at $R\; \in [4,\; 4.5]\; (R_E)$ versus $|\sigma|$ with $\Delta J$ time lagged by $\sim 3$ days to obtain the highest Spearman correlation coefficient (see methods). The orange solid line and the surrounding shaded area are the median value and the interquartile range of the data binned on a discrete grid. Panels g, m, and s are 2D histograms of $\Delta J$ versus $V$, the solar wind bulk speed with the same time lag applied. Finally, panels h, n, and t are 2D histograms of $|\sigma|$ versus $V$. The vertical white lines mark $V = 500$ km/s, approximately separating fast and slow solar wind, and the horizontal white lines mark the median $\Delta J$ (panels g, m, s) and $|\sigma| = 0.4$ (panels h, n, t).  }
    \label{fig_zoomed}
\end{figure}

Fig. \ref{fig_zoomed} zooms in on three intervals covering the solar maximum (Fig. \ref{fig_zoomed}c-h), the declining phase (Fig. \ref{fig_zoomed}i-n), and the solar minimum (Fig. \ref{fig_zoomed}o-t) as highlighted respectively by the red, black, and blue patches in Fig. \ref{fig_zoomed}a. The intervals were selected to minimize the number of ICME events and thereby isolate the coherent variability. In all three intervals, we observe a prominent recurrent flux enhancement at the Carrington periods (Fig. \ref{fig_zoomed}c, i, and o). To identify the source of the solar wind interacting with Earth, we apply a classification technique that assigns probabilities to four possible origins  \cite{xu2015new,camporeale2017classification}.
The four origins are Ejecta (ICME plasma); coronal hole (CH); streamer belt (SB); and sector reversal (SR) solar wind (Fig. \ref{fig_zoomed}d, j, and p). Visually, recurrent black-shaded regions - indicating CH solar wind - coincide with intervals of flux enhancement, particularly during the declining phase (Fig. \ref{fig_zoomed}i, j) and solar minimum (Fig. \ref{fig_zoomed}o, p). By contrast, during solar maximum, the solar wind origin is more complex and not clearly CH-dominated, yet recurrent flux enhancements are still evident (Fig. \ref{fig_zoomed}c,d).

To quantitatively test for the role of Alfv\'enic fluctuations on electron flux, we calculate the absolute value of the normalized cross-helicity $|\sigma|$, which approaches 1 for highly Alfv\'enic wind and zero otherwise (see Methods; Fig. \ref{fig_zoomed}e, k, q). The intervals show a periodic recurrence of high $|\sigma|$ solar wind coincident with the recurrent flux enhancements. During the declining phase and solar minimum, the main source of high Alfv\'enicity solar wind is coronal holes (black patches in Fig. \ref{fig_zoomed}j, p). By contrast, during solar maximum, intervals classified as SB or SR also exhibit high $|\sigma|$ values. This behaviour is consistent with previous observations of Alfv\'enic slow wind originating near coronal hole/active region boundaries (see section \ref{equatorial_coronal_holes} in Extended data, \cite{d2015origin}). For the solar maximum interval, we verified using the Richardson and Cane \cite{richardson2010near} and the Helio4Cast \cite{mostl2026magnetic} ICME lists that only the first flux enhancement (around September 2014) is preceded by an ICME, while the remaining four are associated with slow-Alfv\'enic wind.

Visually, the periodic recurrence of high Alfv\'enicity solar wind coincides with intervals of flux enhancement. To quantify this relationship, we plot the column-normalized probability of the normalized flux $\Delta J$ as a function of $|\sigma|$ for each solar cycle interval (Fig. \ref{fig_zoomed}f, l, r). We account for a $\sim 3$-day time lag that maximizes the Spearman correlation coefficient between $\Delta J$ and $|\sigma|$ (see Methods). Across all solar-cycle phases, we find a consistent positive correlation between $|\sigma|$ and $\Delta J$ (see orange lines in Fig. \ref{fig_zoomed}f, l, r). 

Since fast solar wind is also characterized by high $|\sigma|$ \cite[and references therein]{bruno2013solar,d2021solar}, to decouple the effects of speed and Alfv\'enicity on the flux level we plot $\Delta J$ versus bulk flow speed $V$ and $|\sigma|$ versus $V$ for each of the three intervals in panels (g, h), (m, n), and (s, t) respectively. The vertical line represents $V = 500$  (km/s), which we take to approximately separate fast and slow solar wind \cite{alterman2025characterizing,alterman2025cross}. The horizontal lines in panels g, m, and s are the median value of $\Delta J$ in each of the three intervals, which we use to separate enhanced flux levels within each phase of the solar cycle. As for panels h, n, and t, the horizontal lines represent $|\sigma| = 0.4$. One can directly see that for all three intervals, high solar wind speeds are characterized by high $|\sigma|$ (upper right corner of panels h, n, and t) and a subsequent flux increase (upper right corner of panels g, m, and s). However, what is interesting is the effect of the slow wind on flux. For solar maximum, most of the wind reaching Earth is of the slow-Alfv\'enic type as the distribution in $\left(|\sigma|, \; V\right)$ space is concentrated in the low $V$, high $|\sigma|$ quadrant (panel h). Such slow-Alfv\'enic wind leads to flux enhancement as can be seen qualitatively by comparing panels c, d, and e, and quantitatively in the upper left corner of panel g (also, see section \ref{equatorial_coronal_holes} in Extended data for backmapping the solar source of the solar wind showing it to be coming from an active region-coronal hole boundary, which is consistent with sources of slow-Alfv\'enic wind \cite{d2015origin}). 

These results support a causal link between solar wind Alfv\'enicity and outer radiation belt electron flux enhancement. The Alfv\'en waves embedded in the IMF drive recurrent southward $B_z$ intervals, thereby triggering magnetic reconnection, substorm activity, and ultimately radiation belt flux enhancement, with a $\sim$3 day lag (see section \ref{carrington_scale_flux} in the Extended data, \cite{tsurutani1995interplanetary,lee2006repetitive,tsurutani2006corotating,ala2024impact}). A similar time lag of $\sim 1-3$ days has been reported in the literature \cite[and references therein]{paulikas1979effects,baker1990linear,hajra2014relativistic,georgiou2018ultralow,weston2025vampire}, with the precise mechanisms governing this time lag warranting further investigation.

High solar wind speed is not a necessary condition for flux enhancement, since slow-Alfv\'enic wind also leads to electron energization. This does not, however, exclude a complementary role for solar wind speed in enhancing the efficiency with which Alfv\'enic fluctuations energize radiation belt electrons. The declining phase, characterized by recurrent exposure to coronal hole wind that is both Alfv\'enic and fast, exhibits higher fluxes than the other phases of the solar cycle. This may reflect an additional role for speed, but it can equally be explained by the recurrence itself: repeated exposure replenishes the belt before fluxes can decay to the background levels seen, for example, at solar maximum. Quantifying how much speed adds to the efficiency of Alfv\'enic driving requires a dedicated analysis, which we leave for future work.

Our results can also clarify the physical origin of the well-known triangular, non-linear relationship between electron flux and solar wind speed \cite{paulikas1979effects,reeves2011relationship}. At lower solar wind speeds, the relativistic flux spans a broad range of values, while at higher speeds, the flux is consistently elevated. This is most clearly seen during the solar minimum interval (Fig. \ref{fig_zoomed}s), which spans sufficient duration to sample all three solar wind types at comparable frequencies. Fast solar wind consistently exhibits high $|\sigma|$ (see upper right quadrant of Fig. \ref{fig_zoomed}h, n, and t), and given that high $|\sigma|$ intervals lead to flux enhancement, fast wind will therefore lead to elevated flux levels. Slow solar wind, however, can be either Alfv\'enic or non-Alfv\'enic \cite{d2015origin,d2021solar}. Slow-Alfv\'enic wind drives flux enhancement while slow non-Alfv\'enic wind does not, naturally producing the observed spread in flux values at low speeds and giving rise to the triangular relationship.

\section{Summary and Conclusions}\label{discussion}

\begin{figure}[htbp!]
    \centering
    \includegraphics[width=.6\linewidth]{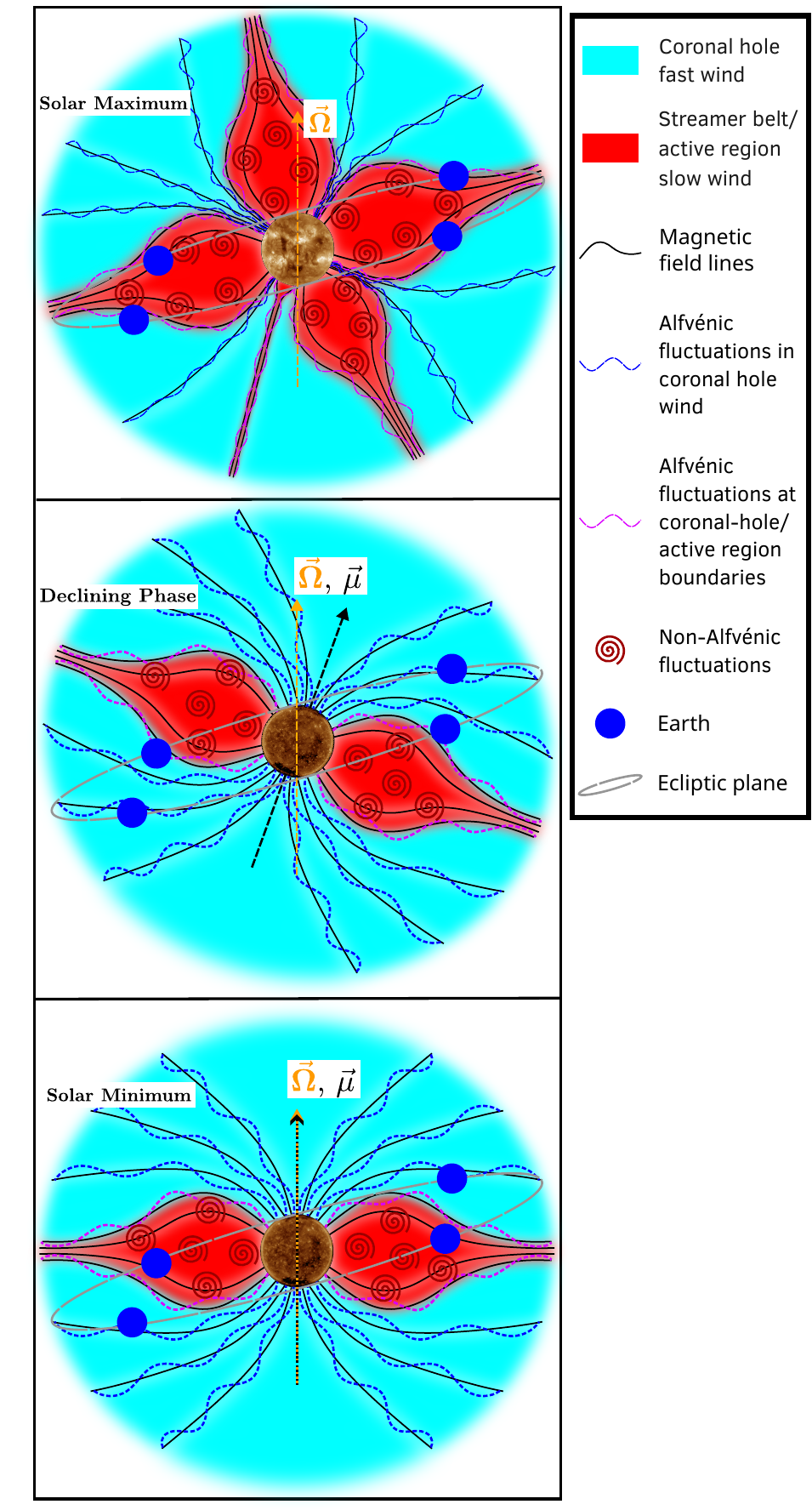}
    \caption{A sketch summarizing the interaction between Earth and the solar magnetic field as it evolves during a solar cycle. The top panel represents the solar maximum, where the dipole moment is negligible and the higher-order moments dominate the magnetic field. During Earth's orbit (blue circle), it mostly sees streamer belt solar wind (shaded in red) with little exposure to coronal hole solar wind (shaded in cyan). The middle panel represents the declining phase, with the dipole moment, $\vec{\mu}$, emerging again and being tilted to the rotational axis of the sun $\vec{\Omega}$. During that phase, Earth is mostly exposed to the solar wind of coronal hole origin. The bottom panel represents the solar minimum, where the dipole moment is mainly aligned with the solar rotational axis, and the Earth is periodically exposed to coronal hole solar wind during the equinoxes (every 6 months) and with every Carrington rotation. The blue oscillating lines overlayed on top of the magnetic field lines in the cyan-shaded regions represent the high Alfv\'enic character of the coronal hole solar wind. The red swirls in the red shaded regions represent the low Alfv\'enic character of the SB solar wind. The magenta oscillating lines at the boundary between the coronal hole and the streamer belt wind represent slow Alfv\'enic wind. All panels show a meridional view of the sun. The images of the sun used have been obtained from the AIA instrument on NASA's Solar Dynamics Observatory (SDO) at 193 Angstrom wavelength at different phases of the solar cycle.}
    \label{fig:sketch}
\end{figure}

In this study, we use seven years of Van Allen Probes relativistic electron flux data, combined with solar wind measurements at Earth's orbit from the OMNI database and synchronic maps of the photospheric magnetic field, to establish a direct link between the solar magnetic configuration and Earth's outer radiation belt.
We challenge the current prevailing paradigm that the seasonal variation is driven by geometric effects (the Russell-McPherron and equinoctial effects) and provide direct evidence that relativistic electron flux is driven by the recurrence of coronal hole solar wind (the axial effect). Given previous work, we unify the mechanism driving coherent outer-belt variability across multiple timescales, from solar-cycle to seasonal and Carrington periods.
Finally, we present strong observational evidence that the Alfv\'enic nature of the solar wind is a key driver of electron energization in the outer radiation belt, and that high solar wind speed is not a necessary condition for this energization.

Fig. \ref{fig:sketch} provides a visual overview of the proposed model. At solar maximum, the solar dipole moment is weakest as it flips polarity, and the magnetic field is dominated by higher-order moments \cite{woo2002origin,livshits2006variations,owens2013heliospheric}. As a result, most of the solar wind reaching Earth originates from the streamer belt, with coronal hole solar wind occurring only sporadically (upper panel of Fig. \ref{fig:sketch}). During this phase, there is no seasonal variation in the electron flux, and the primary driver of electron energization is geomagnetic storms driven by ICMEs with occasional and small-scale equatorial coronal holes (see \ref{equatorial_coronal_holes}, \cite{owens2013heliospheric,richardson2012solar}).

During the declining phase, the solar dipole moment dominates again, tilted with respect to the solar rotational axis. This configuration allows polar coronal holes to extend to lower heliographic latitudes, and the solar wind interacting with Earth would be of CH origin once or twice every Carrington rotation, depending on the exact configuration of the solar magnetic field. Consequently, modulation of the electron flux at the Carrington periods dominates this phase (middle panel of Fig. \ref{fig:sketch}). The sustained periodic energization of the electrons leads to a peak in their flux relative to other phases of the solar cycle.

Finally, during solar minimum, the solar dipole moment is nearly aligned with the solar rotational axis. Due to the tilt between the ecliptic and equatorial planes, Earth's heliographic latitude oscillates from $-7.25^\circ$ to $7.25^\circ$, reaching its maximum near equinoxes. During that period, the probability that Earth is exposed to polar coronal hole solar wind is highest, which leads to an increase in radiation belt electron flux (bottom panel of Fig. \ref{fig:sketch}). Because of the Sun's rotation at the Carrington period, Earth encounters polar coronal holes once or twice per rotation, producing the observed overlap of two periodicities in the electron flux: one at seasonal and the other at Carrington timescales.

The data used in this study are restricted to solar cycle 24; however, we argue that the results presented above generalize to other solar cycles. First, by showing that the slow Alfv\'enic wind drives flux enhancements in addition to those produced by the fast Alfv\'enic wind, we identify Alfv\'enicity as a key driver of electron energization and show that high solar wind speed is not a necessary condition. These conclusions are independent of solar cycle amplitude, as they rest on a physical distinction between wind types rather than on the particular characteristics of cycle 24.
Second, the link between high-speed streams and electron flux enhancement has been documented across multiple solar cycles \cite[e.g.][and references therein]{paulikas1979effects,lyons2005solar,tsurutani2006corotating,reeves2011relationship,li2011behavior,reeves2013electron,hajra2022corotating}; our work refines the physical interpretation of this well-established relationship by identifying Alfv\'enicity as a key driver.
Third, the solar magnetic configuration underlying the model presented in Fig. \ref{fig:sketch} has been shown to hold beyond a single cycle \cite[e.g.][and references therein]{woo2002origin,livshits2006variations,owens2013heliospheric}, and the relative abundance of different solar wind types at 1 AU has been documented across multiple cycles \cite[e.g.][and references therein]{d2015origin,d2021solar}. Together, these studies confirm that the conditions required for the axial effect to drive the seasonal variation are re-established at every solar minimum.

These findings have direct implications for space weather and space safety applications. Satellite mission design already accounts for variations in radiation belt fluxes across the solar cycle, since the total radiation fluence can differ by orders of magnitude between solar maximum and solar minimum, with corresponding consequences for radiation shielding requirements. However, the physical mechanisms driving electron flux enhancements across different phases of the solar cycle have not been fully established. Our results provide a clearer picture of these mechanisms, in particular, the role of high Alfv\'enicity solar wind in driving flux enhancement. This physical understanding can improve the accuracy of predictive models, enabling more accurate forecasting of high-flux periods across multiple temporal scales, from days to years, with direct benefits for satellite safety operations. We are in the process of developing such a physics-informed predictive model.

More broadly, these results have far-reaching implications for understanding the dynamics of Earth's radiation belt and beyond. In particular, we stress the importance of the efficient coupling between the solar wind, which provides the free energy required to accelerate particles to relativistic energies, and the magnetosphere. Due to Earth's northward-directed magnetic field, extended periods of southward IMF are required to be able to couple the solar wind and the magnetosphere and start the energy injection into the system.  During periods of high Alfv\'enicity, such sustained southward IMF intervals naturally arise. We therefore speculate that even for planetary environments with a southward-directed planetary magnetic field, e.g., Jupiter \cite{smith1974magnetic} and Saturn \cite{acuna1980magnetic} or when Earth's magnetic field reverses polarity \cite{valet2005geomagnetic}, high Aflv\'enicity solar wind would still be a strong driver of geomagnetic activity as the Alfv\'enic oscillations generate periods of both northward and southward IMF orientations.

\section{Methods}

\subsection{REPT data processing and calculating $J_{median}$}
From the Van Allen Probes A and B, we use the level 3 (L3) differential electron flux measurement provided from the Relativistic Electron Proton Telescope (REPT) instrument \cite{baker2013relativistic} at the temporal resolution of $\sim 10$ seconds at energies ranging from 1.9 to 20 MeV.
We first average the electron flux measured over the pitch angle to obtain the pitch angle-averaged flux $J(t,\; E)$ for each VAP A and B. We then discretize the data from both probes on a grid of the geocentric radial distance R in units of Earth radii ($R_E$), and obtain the combined flux from VAP A and B as a function of time, geocentric radial distance, and energy. 
The time steps at each bin in R are not uniform due to the elliptic nature of the orbit of VAP A and B. This can be problematic, especially when applying spectral analysis to the data by calculating the Fourier and wavelet transforms. Therefore, to mitigate this problem we upsample all of the data to a time resolution of 1 hour by linearly interpolating between data points (in log space) and combine them in a single time series to obtain  $J(t,\; R,\; E)$ plotted in Fig. \ref{fig_overview}b. This is justified since the timescales of interest ($\sim$ 10s of days and longer) exceed the largest data gap and time step (a few hours) by roughly two orders of magnitude, so interpolation introduces negligible distortion at the periods relevant to our conclusions.

In most previous studies, either McIlwain's L \cite{1961McIlwain} or Roederer's $L^*$ \cite{2014Roederer} parameters are used in the spatial discretization; however, we chose to discretize along the spacecraft geocentric radial distance R instead, since both the L and $L^*$ parameters rely on an assumed geomagnetic field, which might not be accurate at extreme geomagnetic conditions. Furthermore, for the purpose of incorporating our results in a future predictive model, the spacecraft geocentric radial distance is the most relevant quantity for the safety of satellites in orbit. Finally, we have checked that our results and conclusions do not change if we use R or L.

Additionally, to properly quantify the variations in the electron flux, we normalize the flux to the seven-year median flux profile  $\Delta J = \log_{10}\left(\frac{J}{J_{median}}\right)$, which we show in Fig. \ref{fig_overview}c. 

\begin{figure}[htbp!]
    \centering
    \includegraphics[width=.8\linewidth]{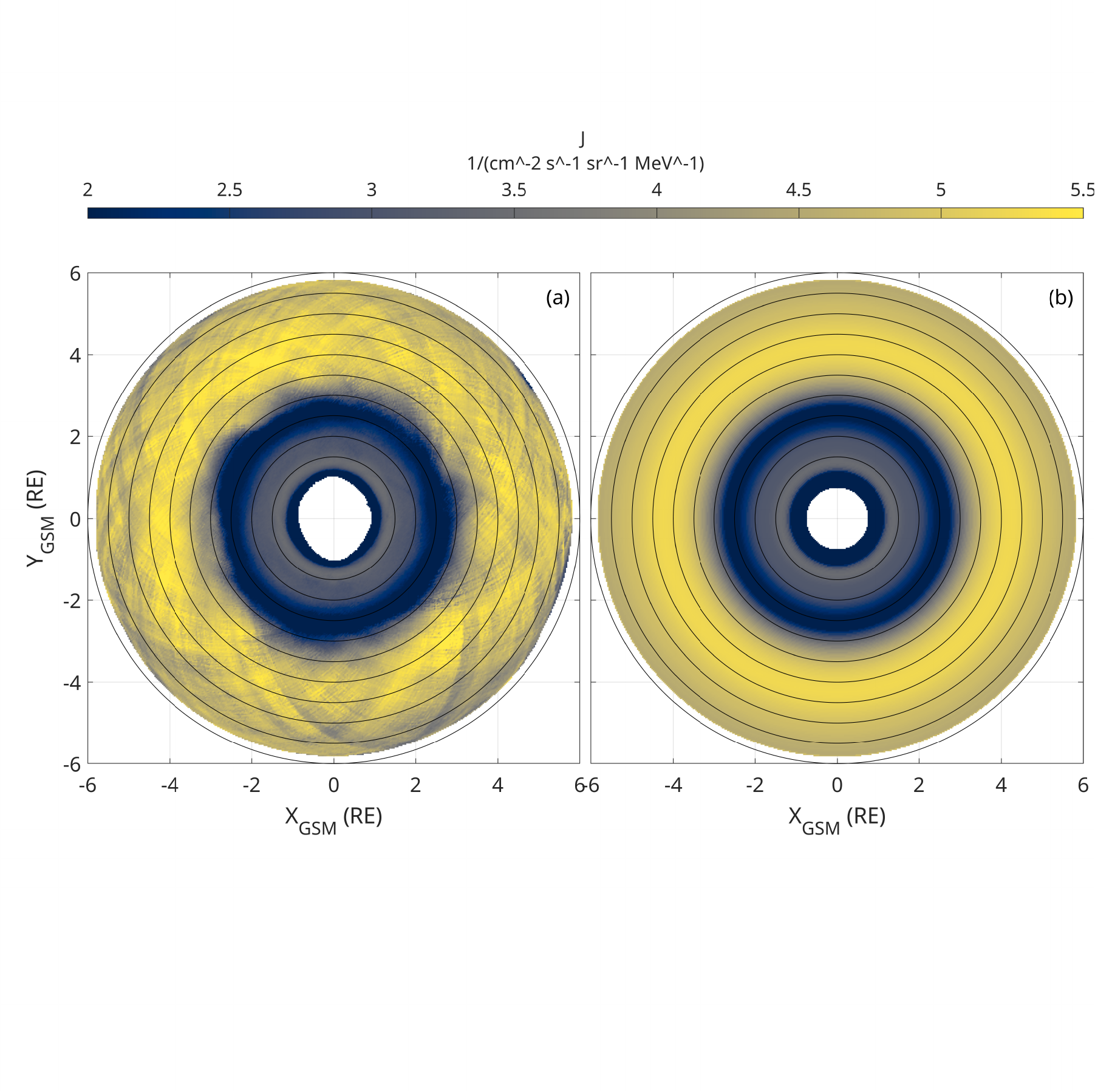}
    \caption{The 2D median flux profile for the 1.9 MeV energy channel. Panel a shows the profile without averaging over the azimuthal direction. Panel b shows the median flux obtained after averaging over the azimuthal direction and is the one referred to as the median flux profile $J_{median}$ in our report.}
    \label{fig_Jomni}
\end{figure}

To compute a median flux profile as a function of energy and geocentric radial distance, we discretize the 3D space around Earth on a GSM-coordinate grid with a grid spacing of 200 km. From both VAP A and B, we have a total of around $\sim 40$ million flux measurements. We bin those measurements on the discrete grid based on the spatial location of the spacecraft. We then calculate the median value of the flux in each bin, and calculate the median along the GSM-Z-axis to obtain a 2D median equatorial flux profile of the electrons as a function of the geocentric radial distance R and energy E  based on seven years of VAP A and B measurements. Fig. \ref{fig_Jomni} a shows this median profile for the 1.9 MeV energy channel. Because of the temporal variability of the radiation belts, the obtained profile is not uniform along the azimuthal direction. To obtain the final model for the electron flux in Earth's radiation belts, we compute the median value within rings of constant R. The resulting profile is shown in panel b, and is the profile of $J_{median}$ used to calculate $\Delta J$.

We observe variability in $J$ and $\Delta J$ at energy channels $<7.7$ MeV. However, in both Figs. \ref{fig_overview} and \ref{fig_zoomed}, we choose to plot only the fluxes at $E = 1.9$ MeV as they show the largest variability and are representative of the coherent dynamics at higher energies; however, our results do not change for different energy channels (see Extended Data).

Finally, it is worth mentioning that in the seven-year median flux profile shown in Fig. \ref{fig_Jomni}b, the peak flux occurs at $R \in [4,\; 4.5] \; R_E$. To capture the maximum variability, we choose this range in R to calculate the wavelet and Fourier transforms shown in Fig. \ref{fig_overview} d and h. We have verified that repeating the analysis for different ranges in $R$ will not affect our main conclusions.

\subsection{Solar wind data and source classification}

We use NASA's OMNI database \cite{king2005solar} to obtain the sunspot number at 1-hour resolution and the various solar wind parameters at 1-minute resolution (\url{https://spdf.gsfc.nasa.gov/pub/data/omni/}). The sunspot number is calculated by the Royal Observatory of Belgium (\url{https://www.sidc.be/}). The solar wind parameters (magnetic field, velocity, density, etc) are measured in situ at the first Lagrange point (L1) in the near-Earth space from multiple satellites (such as ACE and Wind). The measurements are then propagated to the nose of Earth's bow shock.

To identify the source of the solar wind interacting with Earth, we use a classification technique \cite{xu2015new,camporeale2017classification} applied to solar wind properties from the OMNI database \cite{king2005solar}, which returns the probability of the solar wind's origin being one of four different categories: Ejecta (ICME plasma), coronal hole (CH), streamer belt (SB), and sector reversal (SR) solar wind. \cite{camporeale2017classification} demonstrated that the maximum probability at each time step corresponds to the correct solar wind type in more than 90\% of cases, and more than 99\% of cases for coronal hole wind. Since the algorithm outputs a probability --- the quantity plotted in Fig.~\ref{fig_zoomed} --- ambiguous or undecided classifications are straightforward to identify, making this probabilistic approach well-suited to our qualitative assessment of whether the solar wind is of coronal hole origin or not. We corroborate the classification using independent measures. For the solar maximum interval, a backmapping technique confirms that the solar wind emanates from an active region/coronal hole boundary, the known source of slow Alfv\'enic wind (see section \ref{equatorial_coronal_holes} in Extended data, \cite{d2015origin}). During the declining phase and solar minimum, intervals classified as coronal hole origin (black patches in panels j and p of 
Fig.~\ref{fig_zoomed}) also exhibit elevated $|\sigma|$ (panels k and q), and statistically high $|\sigma|$ solar wind is associated with high $V$ (panels n and t) --- both characteristic properties of high-speed streams from coronal holes.

\subsection{Alfv\'enicity of the solar wind}
\begin{figure}[htbp!]
    \centering
    \includegraphics[width=0.8\linewidth]{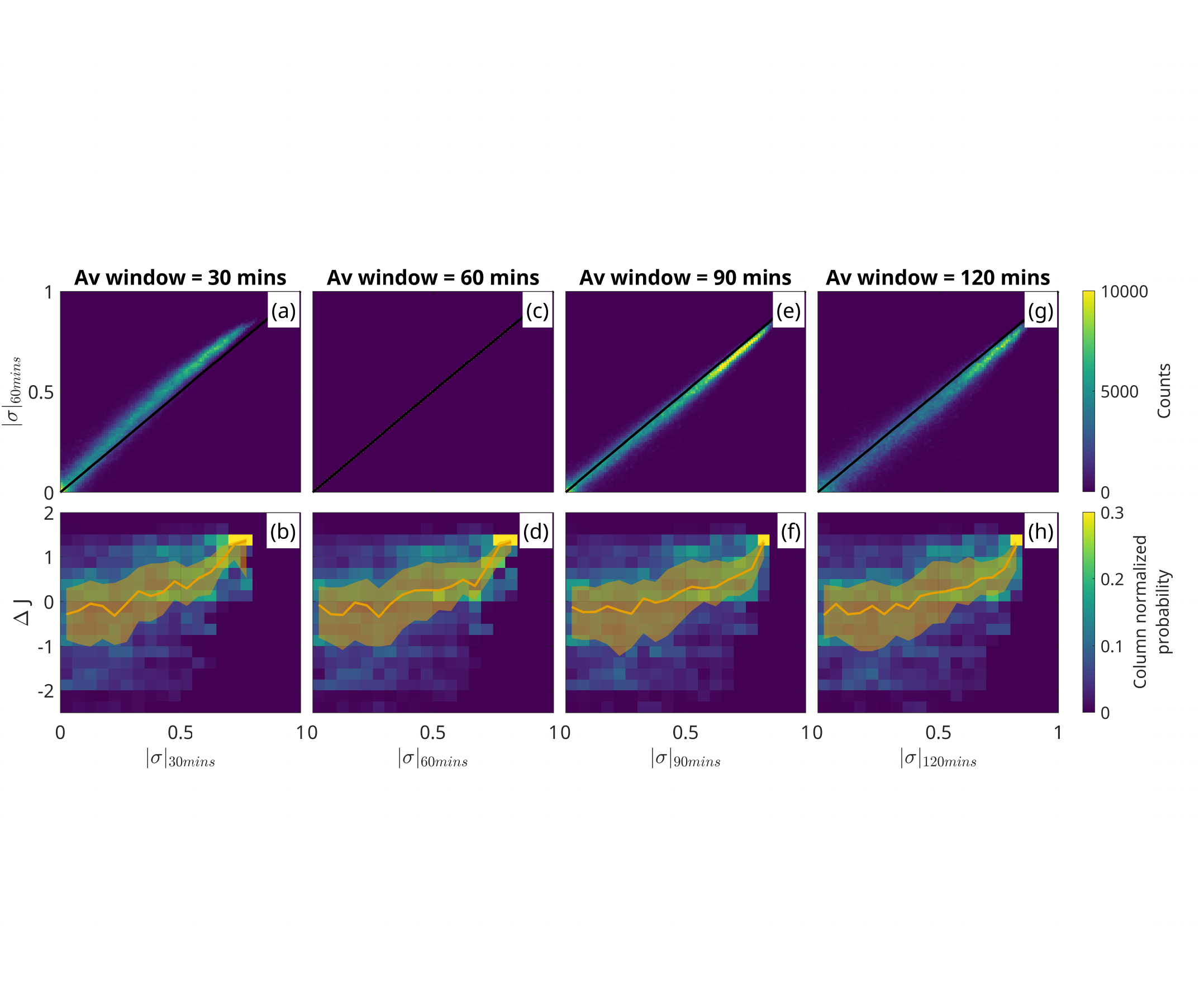}
    \caption{Robustness of results to variations in the averaging window used to compute $|\sigma|$. Each column corresponds to a different averaging window: 30 minutes (a, b), 60 minutes (c, d), 90 minutes (e, f), and 120 minutes (g, h). Note that to produce this plot, we use the 1-day smoothed $\sigma$ rather than the raw values. Top row (a, c, e, g): 2D histograms comparing $|\sigma|$ computed with each window against $|\sigma|$ computed with the 60-minute window, showing a tight linear correlation across all cases with only minor differences in magnitude. Bottom row (b, d, f, h): 2D histograms of $\Delta J$ versus $|\sigma|$ for the solar minimum interval (blue interval in Fig. \ref{fig_zoomed}a), with the running median overlaid in orange. The consistency of both the distribution and the running median across all columns confirms that the positive correlation between $\Delta J$ and $|\sigma|$ is robust to the choice of averaging window.}
    \label{fig:averaging_window_sigma}
\end{figure}

To quantify the degree of Alfv\'enicity of the solar wind we use the normalized cross helicity \cite{roberts1987nature, bruno2013solar, d2021solar} defined as:
\begin{equation}
    \sigma = \frac{2 \delta \mathbf{v} \cdot \delta \mathbf{b}}{ \delta v^2 +\delta b^2},
\end{equation}
where $\delta \mathbf{v} = \mathbf{V} - \mathbf{V}_0$ is the fluctuating velocity with $\mathbf{V}_0$ the average velocity in the analyzed interval. Furthermore, 
\begin{equation}    
\delta \mathbf{b} = \frac{\delta \mathbf{B}\; v_A}{B_0} = \frac{\left(\mathbf{B} - \mathbf{B}_0\right)\; v_A}{B_0},
\end{equation}
is the fluctuating magnetic field in Alfv\'en units, with $v_A$ being the Alfv\'en speed, $\mathbf{B}_0$ the average magnetic field, and $B_0$ its magnitude. To obtain $\mathbf{V}_0$ and $\mathbf{B}_0$ we average over a 1-hour-long window, chosen to capture the Alfv\'enic fluctuations in the solar wind. 
This choice is in accordance with recent works investigating solar wind Alfv\'enic fluctuations \cite[and references therein]{d2015origin,d2021solar,alterman2025cross}. Shorter intervals reduce statistical reliability and begin to undersample the correlation scale, yielding statistically unreliable estimates, while longer intervals mix fluctuations across the inertial and energy-containing ranges and increasingly risk sampling heliospheric sector boundary crossings, which can artificially reduce the magnitude of $\sigma$. The 1-hour window, therefore, represents the established optimal choice for capturing Alfv\'enic fluctuations in the solar wind \cite{bavassano1998cross,wicks2013correlations,isaacs2015systematic}.
We have verified that our results are not significantly affected by varying the averaging window between 30 minutes and 2 hours. Fig. \ref{fig:averaging_window_sigma} shows this comparison, where each column corresponds to a different window length. The top row shows that $|\sigma|$ values obtained from different windows are linearly correlated, with only minor differences in magnitude, while the bottom row reproduces the $\Delta J$ versus $|\sigma|$ scatter plot for the solar minimum interval (blue interval in Fig. \ref{fig_zoomed}a), confirming that the key result is robust across all tested windows. Note that to obtain $|\sigma|$ throughout this work, a 1-day running median smoothing is applied to $\sigma$ before taking the absolute value, which we describe below.

\begin{figure}[htbp!]
    \centering
    \includegraphics[width=1\linewidth]{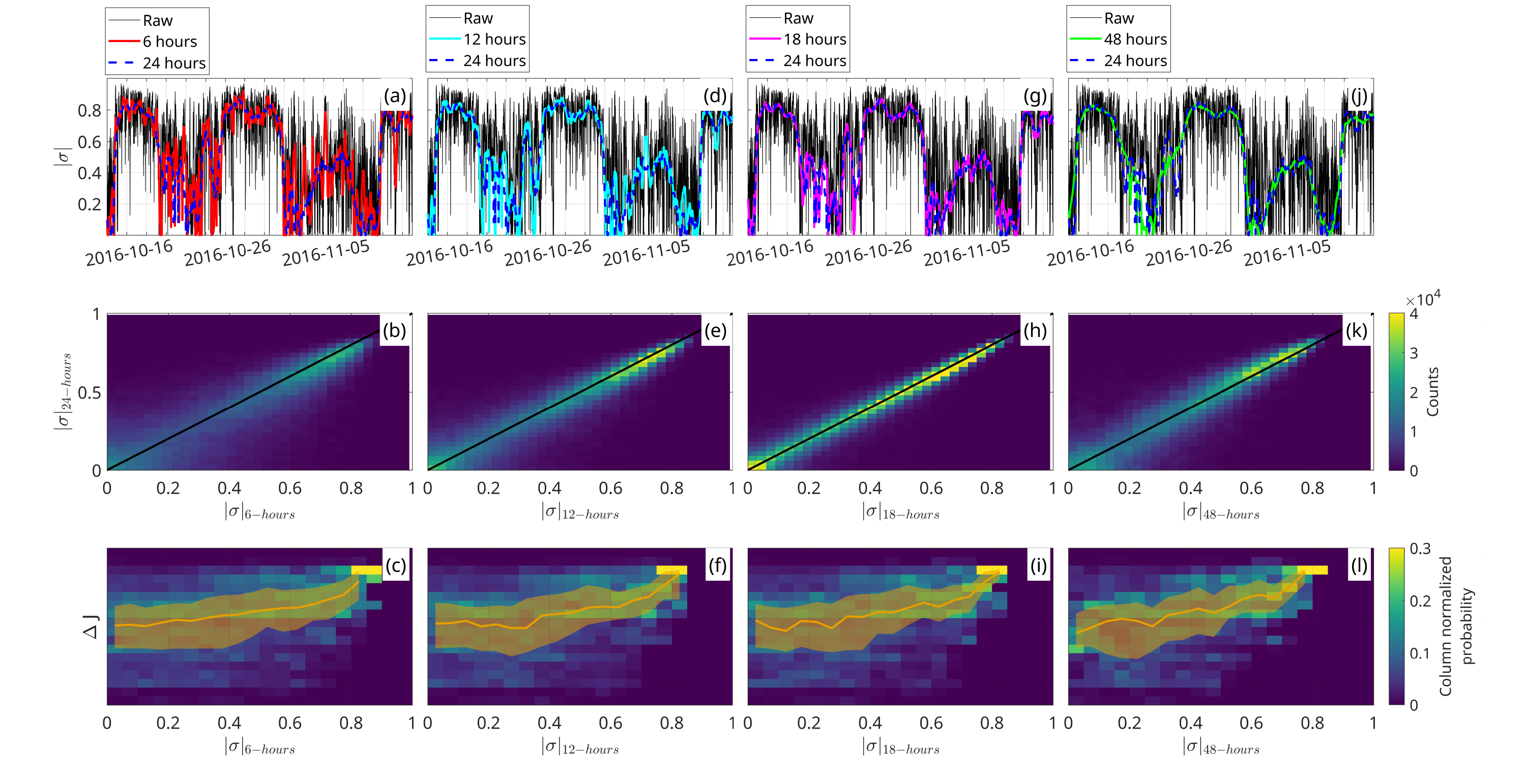}
    \caption{Robustness of results to variations in the smoothing window applied to $\sigma$. Each column corresponds to a different smoothing window: 6 hours (a, b, c), 12 hours (d, e, f), 18 hours (g, h, i), and 48 hours (j, k, l), each compared against the 24-hour smoothing window used in this study. Top row (a, d, g, j): Time series of the raw 1-minute cadence $|\sigma|$ (black), the smoothed $|\sigma|$ for each tested window (colours), and the 24-hour smoothed $|\sigma|$ (dashed - blue), shown for a representative subinterval during the declining phase to better illustrate the effect of the smoothing window on the time series. Middle row (b, e, h, k): 2D histograms comparing $|\sigma|$ smoothed with each tested window against the 24-hour smoothed $|\sigma|$ for the solar minimum interval (blue interval in Fig.~\ref{fig_zoomed}a), with the identity line shown in black. The distributions cluster tightly around the identity line in all cases, with modest deviations for the shortest (6 hours) and longest (48 hours) windows. Bottom row (c, f, i, l): 2D histograms of $\Delta J$ versus $|\sigma|$ for the same interval, with the median profile overlaid in orange. The consistency of the median profile across all columns confirms that the positive correlation between $\Delta J$ and $|\sigma|$ is robust to the choice of smoothing window.}
    \label{fig:smoothing_window_sigma}
\end{figure}

To calculate $\sigma$, we slide the 1-hour averaging window at a 1-minute stride, yielding a time series at a 1-minute cadence. Each value of $\sigma$ reflects the Alfv\'enic correlation of fluctuations within that 1-hour interval.
Since we are interested in the cumulative effect of sustained periods of elevated Alfv\'enicity — on timescales of days to weeks — rather than in short timescale fluctuations, we apply a 1-day moving median to $\sigma$ to suppress higher-frequency variability. This smoothing window is chosen to be long enough to span many periods of the energy-containing Alfv\'enic fluctuations (which have characteristic periods of order 1–few hours at 1 AU \cite{roberts1987nature,bruno2013solar}), thereby capturing the sustained average Alfv\'enic state of the solar wind rather than individual wave packets, while remaining short enough not to affect the longer-term trends we are investigating ($>$10 days). The 1-day and $>$10-day timescales are therefore well-separated in frequency, ensuring the smoothing does not bias our results. 
We verified this by confirming that the key correlations are not sensitive to moderate variations in the smoothing window (between 6 hours and 2 days), as shown in Fig.~\ref{fig:smoothing_window_sigma} for the solar minimum interval (blue interval in Fig.~\ref{fig_zoomed}a).

Finally, the quantity $\sigma$ is equal to -1 or 1 if the solar wind is dominated by Alfv\'en waves propagating in the field-aligned or anti-field-aligned directions, respectively. Since at 1 AU the IMF changes polarity regularly, for every interval analyzed, if $\mathbf{B}_0$ points sunward, we flip it so it's pointing anti-sunward. This way, whenever $\sigma = -1 \; (1)$, it means that the Alfv\'en mode propagates anti-sunward (sunward).

\subsection{Correlation analysis between the flux enhancement and $|\sigma|$}
To quantify the relationship between the flux enhancement $\Delta J$ and the normalized cross helicity $|\sigma|$, we resample both variables onto a common timeline and compute the Spearman correlation coefficient $\rho$, which captures monotonic trends without assuming a linear relationship. After trying multiple time shifts, we find the maximum value of the correlation coefficient to be when $\Delta J$ is lagged by $\sim 3$ days for all three intervals analyzed. The values of $\rho$ for the solar maximum, declining phase, and solar minimum intervals shown in Fig. \ref{fig_zoomed} range between 0.3 and 0.4, with a p-value consistently $\sim0$. This means that we have a moderate and statistically relevant positive correlation between $\Delta J$ and $|\sigma|$ throughout the solar cycle phases. Despite our effort to minimize their numbers, there are still some ICME events in each interval analyzed, which is the cause of the large scatter in the data, particularly for $|\sigma|<0.6$. As expected, this is most prominently seen during solar maximum (Fig. \ref{fig_overview}f). Since the correlation coefficient measures the strength of a monotonic relationship across the full distribution, large scatter in the data will yield a reduced coefficient.

The physical relevance of the trend is, however, clearly visible in Fig. \ref{fig_zoomed}f, l, and r: the median flux (orange line) increases by 1–2 orders of magnitude between low and high $|\sigma|$, a variation that is physically meaningful regardless of the scatter. When the correlation analysis is applied to this median trend alone — effectively filtering the noise — the correlation coefficients rise to $\sim 0.8-0.9$, confirming that the underlying relationship is strong.

\subsection{Surface footpoint connectivity}

To obtain the connectivity footpoints of Earth at the solar surface (see Fig. \ref{fig_overview}g and Fig. \ref{fig_SDO}), we used a two-step ballistic backmapping model. This is a coupled model consisting of the heliospheric and the coronal magnetic field. The first part of the model ballistically backmaps the Earth's position to the outer boundary of the coronal model at a source surface height of 2.5 R$_{\odot}$, where the magnetic field becomes radial. To ballistically backmap the Earth's position, we assume a Parker spiral for the heliospheric magnetic field, whose shape is determined from the radial velocity taken from the OMNI database. The second part of the model traces the coronal magnetic field to the surface at 1 R$_{\odot}$ using a potential field source surface (PFSS) model \cite{schatten1969model,altschuler1969magnetic} using pfsspy package \cite{stansby2020pfsspy}. To reconstruct the coronal magnetic field using the PFSS model, we utilise ADAPT/GONG (Air Force Data Assimilative Photospheric Flux Transport, Global Oscillations Network Group \url{https://gong.nso.edu/adapt/maps/}) synchronic magnetograms as the lower boundary condition for the coronal model. These global magnetic field maps combine the solar photospheric magnetic field along with a surface flux transport model and data assimilation to give the most realistic representation of the Sun's global magnetic field. To select the appropriate ADAPT/GONG map to construct the PFSS model, we took the radial velocity OMNI data on each day at 00 UT and calculated the solar wind propagation time to retrieve the corresponding Sun time.

\begin{figure}[htbp!]
    \centering
    \includegraphics[width=1\linewidth]{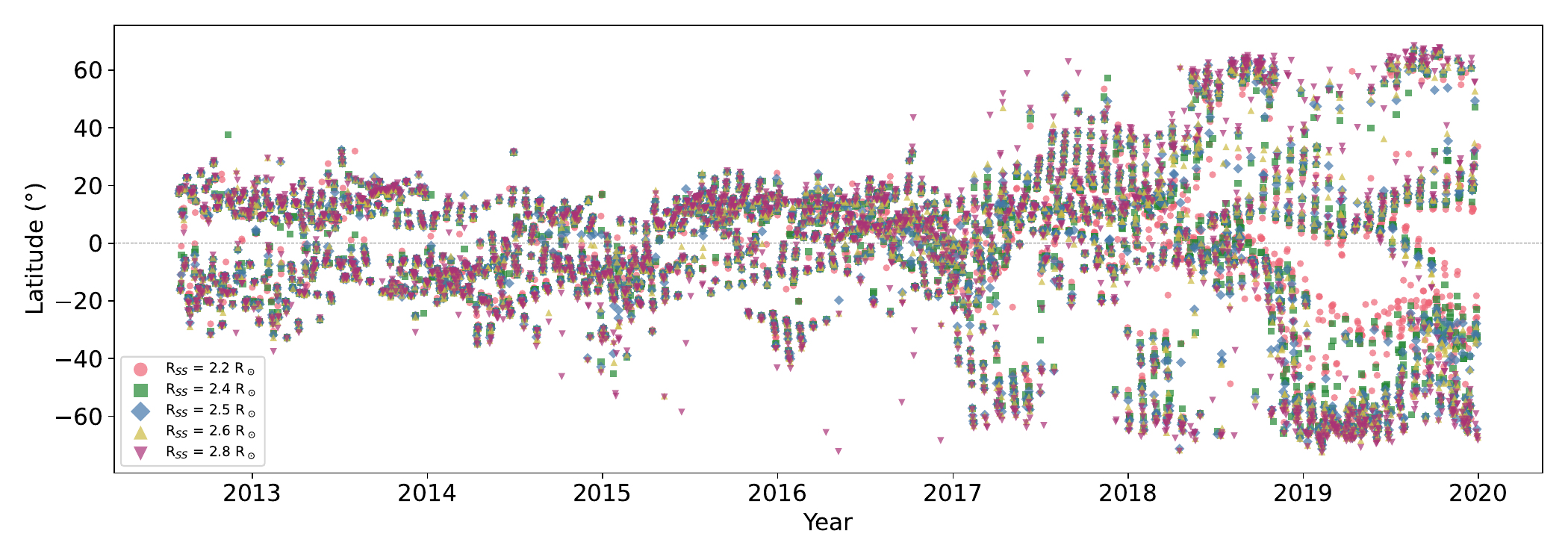}
    \caption{Carrington Latitude of the magnetic connectivity footpoints on the solar surface as a function of time with the source surface heights in the model ranging from 2.2 to 2.8$R_{\odot}$. }
    \label{fig:connectivity_rss}
\end{figure}

Using a PFSS extrapolation for the coronal magnetic field and assuming a Parker spiral for the heliospheric part of the model is a computationally inexpensive but simplified approach. PFSS is a static, current-free model and therefore cannot represent dynamic events or the non-potential nature of the coronal field, which is particularly important for modelling active regions. It also assumes that the coronal magnetic field becomes radial at the source surface height, which may change depending upon the levels of solar activity \cite{riley2006pfss}. However, this approach is reasonable for our purposes since we are interested in Earth's magnetic connectivity over large temporal (a full solar cycle) and global spatial scales.

As the source surface height typically produces the largest uncertainty in the location of the source region of the solar wind \citep{koukras2025uncertainty}, we repeated the analysis varying the source surface height between 2.2 and 2.8$R_{\odot}$ (see Figure~\ref{fig:connectivity_rss}). The results show that the overall trend in the latitudes of the connectivity points with time remains and does not change depending upon the source surface height in the ballistic backmapping model therefore, confirming that our results are not sensitive to the choice of the source surface height.

\section{Data availability }
The data from the Van Allan Probes are publicly available at \url{https://spdf.gsfc.nasa.gov/pub/data/rbsp}. Solar wind parameters and geomagnetic indices from the OMNI dataset are available at \url{https://spdf.gsfc.nasa.gov/pub/data/omni/}. The ADAPT/GONG synchronic magnetograms used for the PFSS modeling are available at \url{https://gong.nso.edu/adapt/maps/}. All post-processed data are available from the corresponding author upon request.
\section{Code availability}
The data was downloaded using the PySpedas Python package documented and available at \url{https://github.com/spedas/pyspedas}. The SciPy Python package \cite{2020SciPy-NMeth} available at \url{https://github.com/scipy/scipy} was used for data processing. The pfsspy package \cite{stansby2020pfsspy} used for the PFSS modelling is available at \url{https://pfsspy.readthedocs.io/en/latest/}. The IRFU-Matlab package \cite{khotyaintsev_2024_14525047} available at \url{https://github.com/irfu/irfu-matlab} was used for data processing and plotting.
\section{Author contribution}
A.L., J.R., and C.W. conceptualized the study and initialized the work. A.L., J.R., C.W., S.Y., and S.R. contributed to the methodology of the work. The analysis was done by A.L., J.R., C.W., S.Y., and S.R. Figures were generated by A.L., while visualization input was provided by A.L., J.R., C.W., S.Y., and S.R. The original draft was written by A.L. with input from all the authors.
\backmatter


\bmhead{Acknowledgements}

We thank the Van Allen Probes team and instrument PIs for data access and support. This work is supported by the STFC grant ST/X001008/1. SLY would like to thank the Science Technology and Facilities Council for the award of an Ernest Rutherford Fellowship (ST/X003787/1). SR acknowledges funding from the Johns Hopkins University Applied Physics Laboratory independent R\&D fund and by NASA DRIVE Science Center for Geospace Storms (CGS) under award 80NSSC22M0163.




\begin{appendices}
\newpage
\section{Extended Data}

\subsection{The equatorial coronal holes in 2013 and 2014}
\label{equatorial_coronal_holes}
\begin{figure}[htbp!]
    \centering
    \includegraphics[width=.8\linewidth]{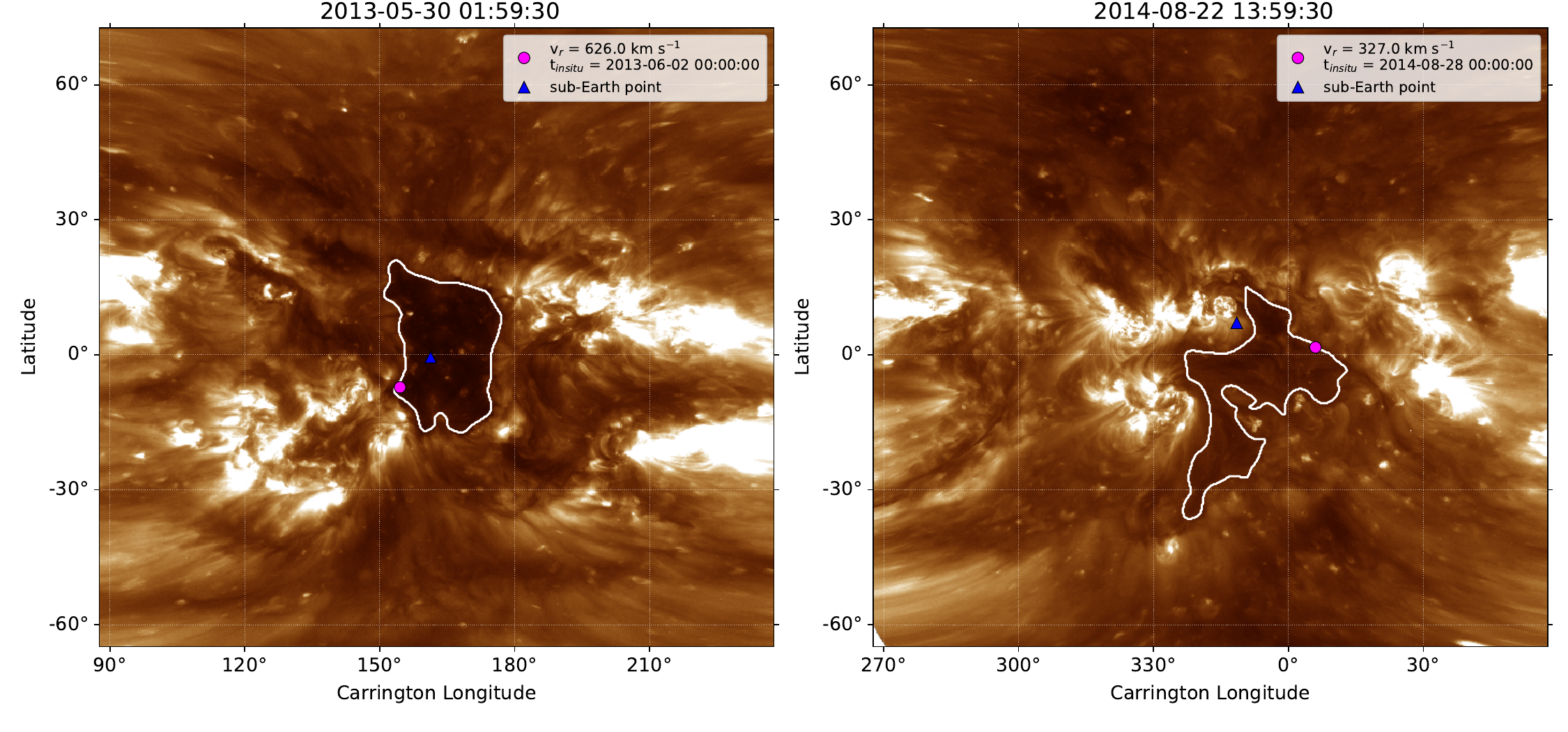}
    \caption{{The equatorial coronal holes during 2013 and 2014 that both caused recurrent flux enhancements. The 193~\AA\ images taken by the Atmospheric Imaging Assembly instrument on board NASA's Solar Dynamics Observatory that have been reprojected into Carrington heliographic coordinates. The dark regions are the equatorial coronal holes seen on 2013-05-30 and 2014-08-22, with the white countours indicating the approximate boundaries of the coronal holes, determined from a threshold applied to the smoothed AIA images. The blue triangle represents the position of Earth backmapped to 2.5$R_{\odot}$ for the time and velocity given in the legend, while the pink circle represents the connectivity footpoints at the surface (1 R$_{\odot}$).}}
    \label{fig_SDO}
\end{figure}

Figure \ref{fig_SDO} shows the solar source of the solar wind reaching Earth on two different dates during solar maximum. Both dates are chosen such that recurrent flux enhancements at the Carrington/sub-Carrington period are observed (see Fig. \ref{fig_overview} and section \ref{carrington_scale_flux}). The white boundary in both panels represents the approximate boundaries of the equatorial coronal holes, while the pink circles represent the back-mapped source of the solar wind at each time. In both cases, the solar wind maps to the boundary between the equatorial coronal hole and  surrounding bright active region(s), which supplements the observation made in the text that the solar wind reaching Earth is of the slow-Alfv\'enic type \cite{d2015origin}.


\subsection{Electron flux at different energies}

\begin{figure}[htbp!]
    \centering
    \includegraphics[width=1\linewidth]{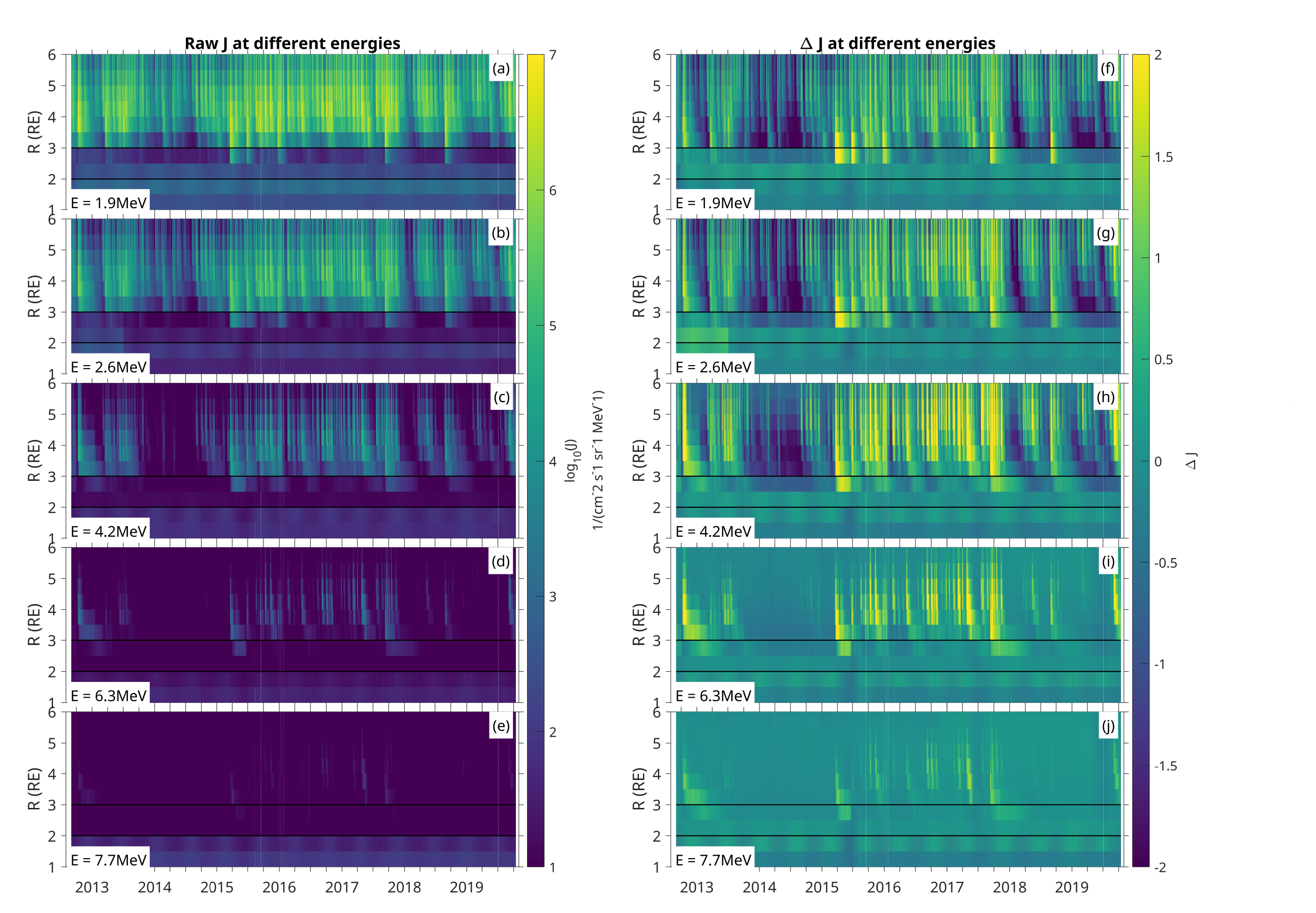}
    \caption{Electron flux and normalized flux across different energy channels. Left panels (a–e) show the raw omnidirectional differential electron flux $log_{10}(J)$ as a function of radial distance R (in Earth radii, $R_E$) and time from 2012 to 2019, at five discrete energies: 1.9, 2.6, 4.2, 6.3, and 7.7 MeV. Right panels (f–j) show the corresponding normalized flux $\Delta J$ at the same energies.}
    \label{fig:energies_j_dj}
\end{figure}

Although we analyze the $1.9$ MeV energy channel throughout the text, the periodic variations that we observe occur at higher energies as well. Fig. \ref{fig:energies_j_dj} shows the full 7 years of RBSP flux $J$ (left column) and normalized flux $\Delta J$ (right columns) for different energies. Periodic flux enhancements at both the Carrington and seasonal timescales are present for energies up to $4.2$ MeV. At higher energies, we still have occasional flux enhancement up to $7.7$ MeV, but not all events energize electrons to such energies.

\subsection{Carrington-scale flux enhancements}

\label{carrington_scale_flux}
\begin{figure}[htbp!]
    \centering
    \includegraphics[width=.8\linewidth]{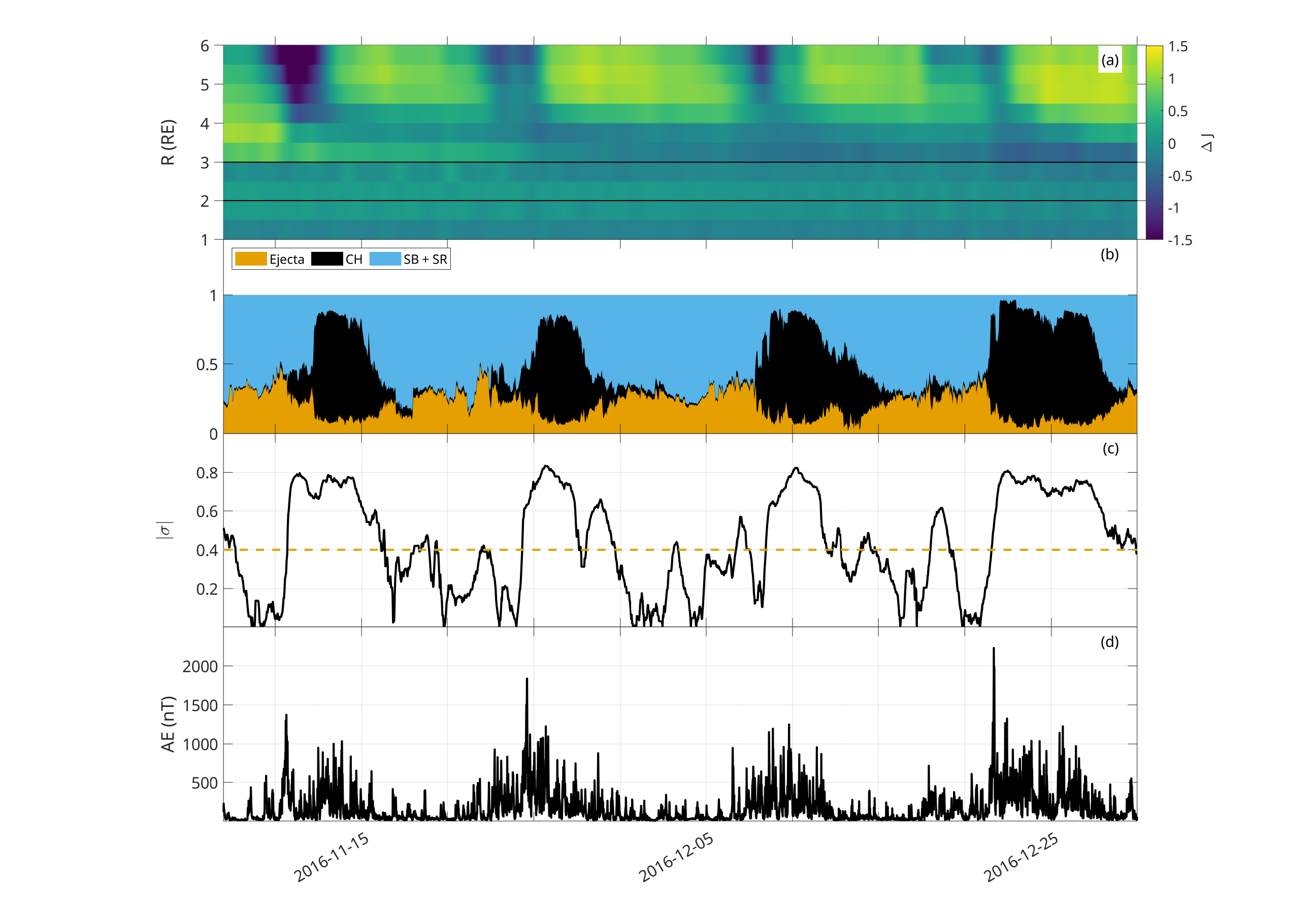}
    \caption{Flux enhancement on the Carrington periods. Panel a shows the time series of $\Delta J(R, E = 1.9 \; MeV, t)$. Panel b shows the same solar wind classification shown in Fig. \ref{fig_zoomed}. Panel c is the absolute value of the normalized cross helicity $|\sigma|$. Panel d is the Auroral Electrojet (AE) index.}
    \label{fig_hildcaa}
\end{figure}

\begin{figure}[htbp!]
    \centering
    \includegraphics[width=1\linewidth]{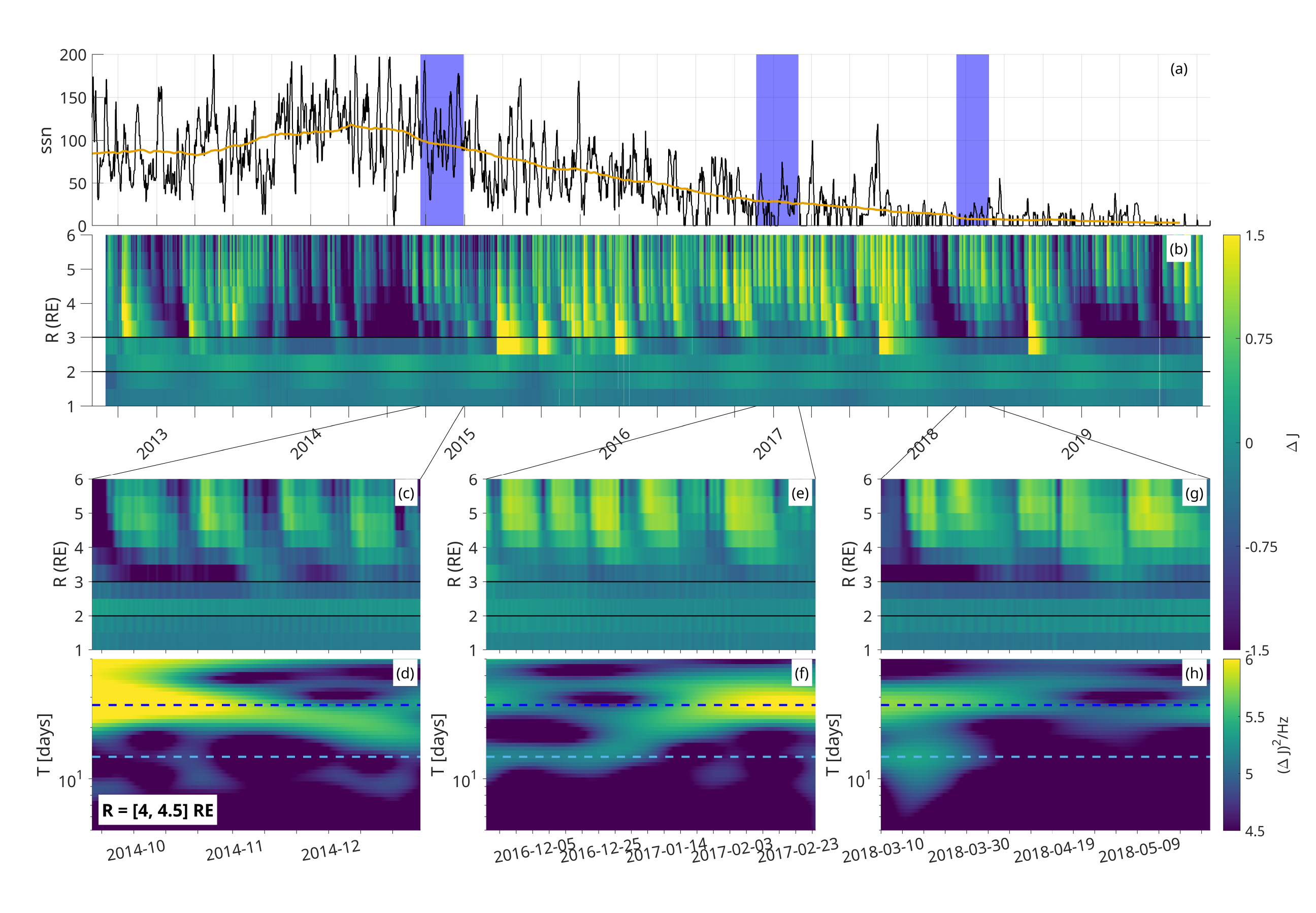}
    \caption{Evolution of the Carrington-scale periodicity in electron flux deviations throughout Solar Cycle 24. (a) Hourly sunspot number (ssn; black) and 1-year running average (orange) over the seven-year Van Allen Probes (VAP) mission, illustrating the progression from solar maximum through the declining phase to solar minimum. Blue shaded regions indicate the three intervals examined in detail below. (b) Normalized electron flux $\Delta J$ at E = 1.9 MeV as a function of radial distance R ($R_E$) and time, with horizontal black lines marking key radial boundaries. Recurrent flux enhancements are visible throughout the mission, modulated on approximately Carrington rotation timescales. (c, e, g) Zoomed-in views of $\Delta J$ for intervals representative of solar maximum, the declining phase, and solar minimum, respectively. (d, f, h) Corresponding wavelet transform power spectral density of $\Delta J$ for each interval. Dark blue and cyan dashed lines mark the Carrington (27 days) and half-Carrington (13.5 days) periodicities, respectively, both of which appear as prominent spectral features, demonstrating that solar rotation-driven modulation of the outer radiation belt persists across different phases of the solar cycle.}
    \label{fig_carrington}
\end{figure}

As proposed in the text, the enhanced Alfv\'enicity of the solar wind causes increased energy input to the magnetosphere, leading to extended substorm activity as quantified by the enhancement in the Auroral Electrojet (AE) index (Fig. \ref{fig_hildcaa}d). This enhancement of AE is nearly simultaneous with the arrival of the high Alfv\'enicity coronal hole solar wind. After a $\sim$ 3-day delay, the electron flux becomes enhanced due to the enhanced substorm activity. Additionally, the recurrence of such flux enhancement events changes dynamically from a $\sim 27$ days to a $\sim 13$ days period depending on solar magnetic configuration as seen in Fig. \ref{fig_carrington}, which zooms in on three intervals from solar maximum, the declining phase, and solar minimum, respectively showing such Carrington-scale periodic activity. Detailed analysis of the dynamics of the radiation belts and their evolution with the evolution of the solar magnetic configuration at such scales is the subject of an ongoing follow-up study.

\subsection{Correlation between the latitude of the backmapped footpoint and the PSD around the seasonal period}
\label{correlation_lat_seasonal}
\begin{figure}[htbp!]
    \centering
    \includegraphics[width=1\linewidth]{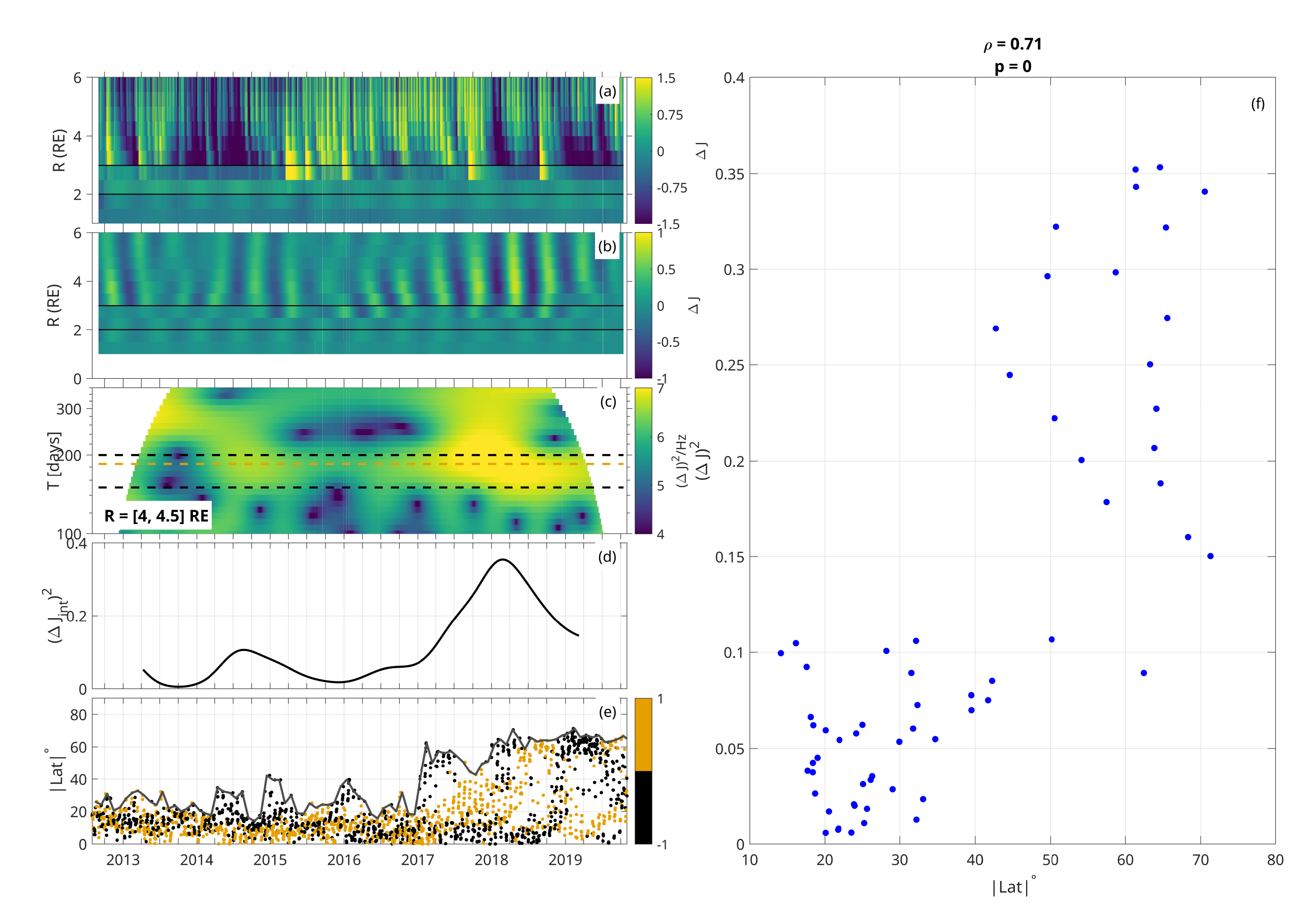}
    \caption{Quantifying the correlation between the seasonal variation in electron flux and the latitude of the solar footpoint of the solar wind. Panel a shows the normalized flux $\Delta J$ as a function of $R$ and time. Panel b shows $\Delta J(R,t, E = 1.9 \; MeV)$ band pass filtered around the period of 185 days (filter interval $[150,\; 200]$ days). Panel c shows the wavelet transform for a cut of $\Delta J$ taken at $R = [4, \; 4.5] \; RE$. Panel d shows $(\Delta J_{int})^2$, the integrated power spectral density around the seasonal period in the interval $[130\; 230]$ days (between the two black dashed lines in panel c ). Panel e shows the absolute value of the latitude at the solar surface of the footpoint of the solar wind reaching Earth. Panel f shows the scatter plot between $(\Delta J_{int})^2$ and the outer envelope of |Lat| (black solid line in panel d).}
    \label{fig_correlat_footpoint}
\end{figure}

To quantify the relationship between the solar wind source footpoint latitude and the coherence of the seasonal variation in electron flux, we show in Fig. \ref{fig_correlat_footpoint}d $(\Delta J_{int})^2$, the integrated power spectral density around the seasonal period in the interval $[130\; 230]$ days (between the two black dashed lines in Fig. \ref{fig_correlat_footpoint}c ). The scatter plot of $(\Delta J_{int})^2$ versus $|Lat|$, the absolute value of the solar footpoint latitude (black solid line in panel e), shows a clear positive correlation with a Spearman correlation coefficient of 0.71 and a p-value $\sim 0$.

\subsection{Visualizing the flux-speed-cross helicity relationship}
\begin{figure}
    \centering
    \includegraphics[width=1\linewidth]{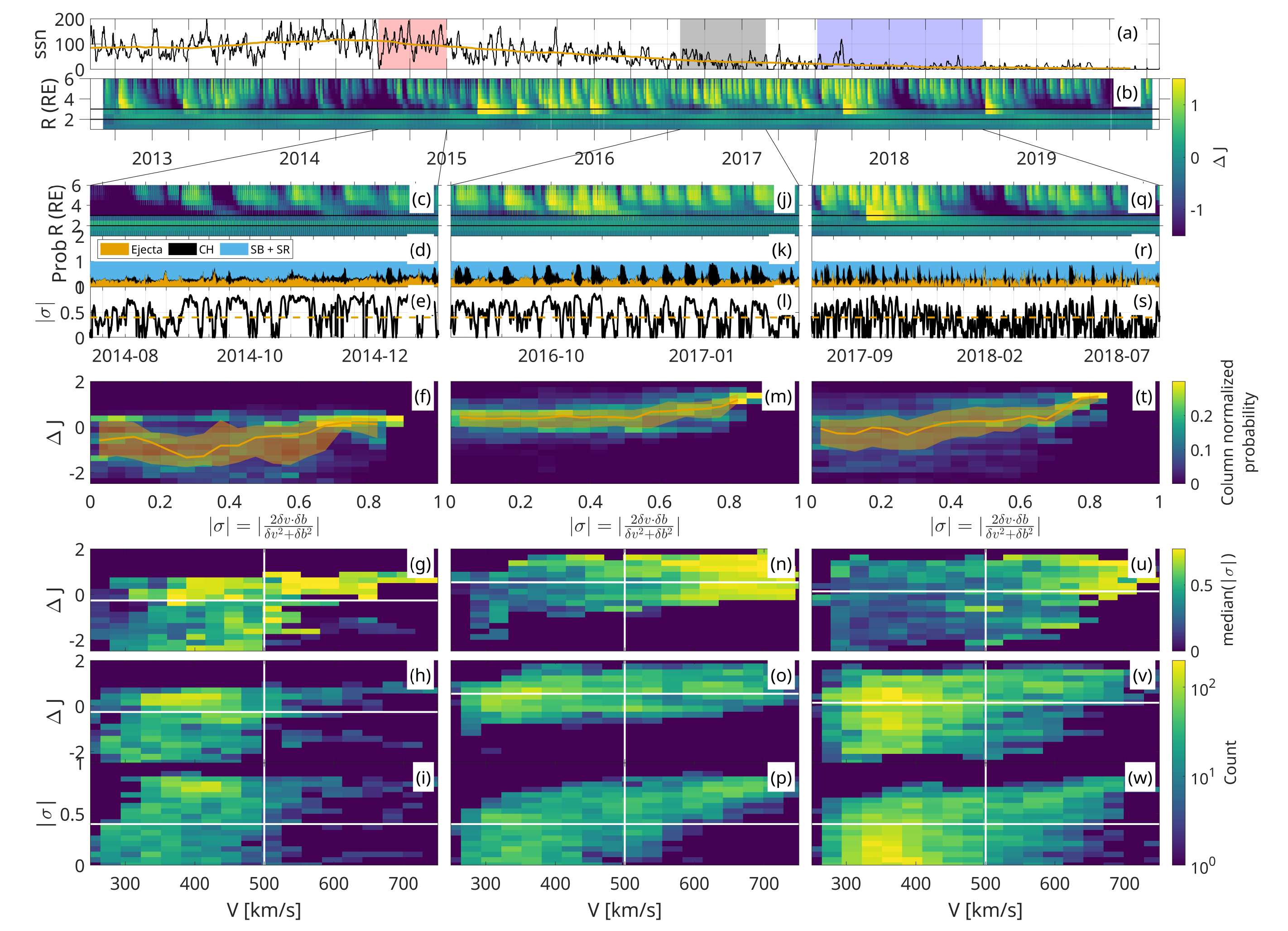}
    \caption{
    Same as Fig.~\ref{fig_zoomed}, with an additional row of panels showing $\Delta J$ versus $V$ colour-coded by the solar wind cross helicity. Panels a-f, j-m, and q-t correspond to Fig.~\ref{fig_zoomed}a-f, i-l, and o-r, and panels i, p, and w to Fig.~\ref{fig_zoomed}h, n, and t. Panels g, n, and u are new: they show the 2D distribution of $\Delta J$ versus the solar wind bulk speed $V$, colour-coded by the median value of $|\sigma|$ in each bin. Panels h, o, and v show the same $\Delta J$--$V$ distribution as 2D histograms of counts, as in Fig.~\ref{fig_zoomed}g, m, and s. Since the colour in panels g, n, and u shows the median of each bin regardless of the number of points it contains, these panels should be read together with the counts in panels h, o, and v. The vertical and horizontal white lines are as in Fig.~\ref{fig_zoomed}.
    }
    \label{fig:fig2zoom_reproduced}
\end{figure}

To help visualize how flux enhancement depends on solar wind speed and Alfv\'enicity, Fig. \ref{fig:fig2zoom_reproduced} reproduces Fig. \ref{fig_zoomed} and adds, above each $\Delta J$ vs $V$ histogram (panels h, o, v of Fig. \ref{fig:fig2zoom_reproduced}), the same histogram colour-coded by the median $|\sigma|$ in each bin (panels g, n, u of Fig. \ref{fig:fig2zoom_reproduced}). During solar maximum, the bins with $\Delta J$ above the interval median include a population at $V < 500$ (km/s) with high median $|\sigma|$, which supports the role of slow Alfv\'enic wind in flux enhancement. Fast wind also lies in the high-$|\sigma|$, high-$\Delta J$ region, consistent with our statement that high speed is not necessary for enhancement while not excluding a role for speed. Since the colour shows the median of each bin, panels g, n, and u should be read together with the counts in panels h, o, and v of Fig. \ref{fig:fig2zoom_reproduced}.

\end{appendices}

\clearpage
\bibliography{sn-bibliography}

\end{document}